\documentclass[reqno]{amsart}
\usepackage{graphicx,amsfonts}

\newtheorem{theorem}{Theorem}[section]
\newtheorem{lemma}[theorem]{Lemma}

\newtheorem{proposition}[theorem]{Proposition}

\newtheorem{remark}[theorem]{Remark}

\newcommand{\reff}[1]{(\ref{#1})}

\newcommand{\cO}{{\mathcal O}}
\newcommand{\cU}{{\mathcal U}}
\newcommand{\cL}{{\mathcal L}}
\newcommand{\cE}{{\mathcal E}}

\newcommand{\cV}{{\mathcal V}}

\newcommand{\pp}{{\partial_{\phi}}}

\newcommand{\tm}{{\frac{t-x}{\epsilon}}}

\newcommand{\T}{{\epsilon t}}

\newcommand{\tu}{{\tilde{u}}}
\newcommand{\ts}{{\tilde{s}}}

\newcommand{\fs}[1]{\ensuremath{\hat{#1}^s}}
\newcommand{\fc}[1]{\ensuremath{\hat{#1}^c}}

\newcommand{\intt}{\int_{0}^{\frac{t}{\epsilon}}}
\newcommand{\argt}{{\frac{t}{\epsilon}-\sigma}}

\newcommand{\ome}{{\omega_{\epsilon}}}

\newcommand{\real}{{\mathbb R}}

\def\bb{\beta}
\def\oo{\omega}
\def\ppy{\partial}

\def\eps{\epsilon}

\begin{document}

\bibliographystyle{unsrt}

\author[Y.Chung]{Y.Chung}
\address[Y.Chung]{Theoretical Division \\ 
                Los Alamos National Laboratory \\ 
                Los Alamos, NM 87545}
\email{ychung@cnls.lanl.gov}
\author[C. K. R. T. Jones]{C. K. R. T. Jones}
\address[C. K. R. T. Jones]{Department of Mathematics\\
         University of North Carolina\\
        Chapel Hill, NC 27599}
\email{ckrtj@email.unc.edu}
\author[T. Sch{\"a}fer]{T. Sch{\"a}fer}
\address[T. Sch{\"a}fer]{Division of Applied Mathematics\\
          Brown University \\
          Providence, RI 02912}
\email{tobias@cfm.brown.edu}
\author[C. E. Wayne]{C. E. Wayne}
\address[C. E. Wayne]{Department of Mathematics and Center for BioDynamics\\
             Boston University\\
             Boston, MA 02215}
\email{cew@math.bu.edu}

\title{Ultra-short pulses in linear and nonlinear media}

\begin{abstract}
We consider the evolution of ultra-short optical pulses in linear and
nonlinear media. For the linear case, we first show that the
initial-boundary value problem for Maxwell's equations in which a
pulse is injected into a quiescent medium at the left endpoint can be
approximated by a linear wave equation which can then be further
reduced to the linear short-pulse equation. A rigorous proof is given
that the solution of the short pulse equation stays close to the
solutions of the original wave equation over the time scales expected
from the multiple scales derivation of the short pulse equation. For
the nonlinear case we compare the predictions of the traditional
nonlinear Schr\"odinger equation (NLSE) approximation which those of
the short pulse equation (SPE). We show that both equations can be
derived from Maxwell's equations using the renormalization group
method, thus bringing out the contrasting scales. The numerical
comparison of both equations to Maxwell's equations shows clearly that
as the pulse length shortens, the NLSE approximation becomes steadily
less accurate while the short pulse equation provides a better and
better approximation.
\end{abstract}

\maketitle

\section{Introduction}

The standard model for describing propagation of pulses in nonlinear
Maxwell's equations is the cubic nonlinear Schr\"odinger equation
(NLSE) \cite{hasegawa-kodama:1995,agrawal:1995}. Two main assumptions
are made in the derivation of the NLSE from Maxwell's equations:
First, it is assumed that the response of the material attains a
quasi-steady-state and second that the pulse width is large in comparison to
the oscillations of the carrier frequency \cite{newell-moloney:1992}.

In most of the applications of the NLSE in the past, i.e., in the case
of pulse propagation in optical fibers, both assumptions were well
satisfied. At present, however, technology for creating very short
pulses has advanced a lot and experiments with pulses which are as
short as a few cycles of the carrier wave have become
possible \cite{karasawa-nakamura-etal:2001}. The description of those
pulses lies beyond the slowly varying envelope approximation leading
to the NLSE \cite{rothenberg:1992}.  Various approaches have
been proposed to replace the NLSE -- see, for example
\cite{blow-wood:1989,mamyshev-chernikov:1990,brabec-krausz:1997}
for a sample of these methods.  In \cite{schaefer-wayne:2003},
building on work of \cite{alterman-rauch:2000},
two of us proposed an alternative model to approximate the
evolution of very short pulses in nonlinear media.

In the present paper we study further the short pulse equation (SPE)
derived in \cite{schaefer-wayne:2003}. There are two main sections of
this paper.  In the first we concentrate on giving a rigorous
justification of several of the assumptions and approximations made in
\cite{schaefer-wayne:2003}, for the {\em linearized} short pulse
equation.  Among the approximations used in \cite{schaefer-wayne:2003}
were:
\begin{enumerate}
\item[(i)] The linearized polarizability of the medium could be approximated
by the Fourier transformed expression
\begin{equation}\label{chi_th}
\hat{\chi}(\omega) = 
c_{\chi} \sum_n |\mu_n|^2 \left\{ \frac{2 \omega_{na}}{(\omega_{na}^2
- \omega^2) + \gamma_{na}^2 - 2i \gamma_{na} \omega} \right\}\ .
\end{equation}
(see equation (3) of Ref. \cite{schaefer-wayne:2003})  Typically,
in deriving \reff{chi_th} one assumes that the medium has reached
some quasi-stationary state, and as J.~Rauch pointed out to us,
it is not clear that for these very short pulses the medium will
have time to reach such a state before the pulse passes.
\item[(ii)] If the expression \reff{chi_th} is an accurate approximation
to the polarizability of the medium, then can one really approximate
solutions of the resulting equation which correspond to ``short''
pulses by solutions of the ``short pulse'' equation derived
in \cite{schaefer-wayne:2003} and if  so, over what time interval
does it provide an accurate approximation to the true evolution?
\end{enumerate}

Note that the first of these points is a question just about the linear
problem and thus in the next section we consider points (i)
and (ii) in the context of a medium whose polarization is assumed
to be linear.  We study solutions of an initial-boundary value
problem in which a linear wave equation is coupled to a medium
whose polarization is modeled by a damped, linear oscillator.
We inject a (short) pulse into the left end of this material
and study how that pulse evolves with time.  We prove rigorously
that one can, even in this short pulse regime, approximate
the polization of the material by the quasi-stationary approximation
\reff{chi_th}, and we show that if the pulse length is measured
by the small parameter $\epsilon$, then the linearized version of
the short pulse equation accurately describes the true solution of
the equation over time scales of ${\cO}(1/\epsilon)$.

The second part of the paper studies of the effect of pulse length on
the propagation of pulses in nonlinear materials. Using the
renormalization group method, we derive both the NLSE and the SPE from
Maxwell's equations. Then we compare the (numerical) evolution of
solutions of a one-dimensional, nonlinear version of Maxwell's
equations with approximations given both by the NLSE and the short
pulse equation. We find as expected from the assumptions that underlie
the formal derivation of these two equations that for slowly modulated
pulses the NLSE does a better job of approximating the evolution, but
as the pulses become shorter and shorter the NLSE approximation breaks
down and the SPE provides a better approximation to the true dynamics.

\section{Formulation of the linear initial-boundary value problem}

In this section we study the propagation of a short pulse
injected at one end of a semiinfinite fiber with
linear polarizability.  If we assume that the polarization
of the electric field is transverse to the fiber then the magnitude
of the electric field, $u(x,t)$, in appropriately
non-dimensional units satisfies the partial differential equation
\begin{equation} \label{u}
u_{xx}=u_{tt}+p_{tt}
\end{equation}
where $p(x,t)$ stands for the (magnitude of the) polarization
of the material, which we assume is parallel to the electric field.

We model the polarization of the material by a damped harmonic
oscillator, so $p(x,t)$ satisfies the equation
\begin{equation} \label{pol_lin}
p_{tt}+\Gamma p_t + \omega_0^2 p = \chi_0 u.
\end{equation}

\begin{remark} In general, the polarization of the medium is
modeled as a sum of oscillators, each with its own natural frequency
and damping constant.  However, as was remarked in \cite{schaefer-wayne:2003},
for infrared pulses in silica fibers, the polarizability 
can be accurately modeled with a single resonance.  Furthermore,
in the present circumstances, the linear nature of the problem
means that if the polarization was modelled as a sum of
several oscillators, we could write the solution as a sum of the solutions
to problems of the type \reff{pol_lin} with different
resonant frequencies and damping constants.
\end{remark}

\begin{remark} Experimentally, the damping of these oscillators
is quite weak -- $\Gamma$ is small.  Thus, throughout this
paper we will assume that $\Gamma^2 - 4 \omega_0^2 < 0$ and
also that $\chi_0 > 0$.
\end{remark}

Physically, we are interested in the situation where
our medium is semi-infinite with one end at the origin and all
fields in the medium are initially zero.  We enforce this
condition by taking initial conditions
\begin{equation} \label{init_lin}
u(x,0)=u_t(x,0)=p(x,0)=p_t(x,0)= 0\;\;{\mathrm{for}}\;\; x>0
\end{equation}
We inject a pulse into the left end of the fiber and model
this by assuming a boundary condition at $x=0$ of the
form
\begin{equation} \label{bound_cond_lin}
u(0,t)=U_0(t)
\end{equation}
where $U_0(t)$ represents the optical pulse.  Since we are interested
in short pulses, we will assume that the injected pulse has the form
$U_0(t) = {\cU}_0(t/\epsilon)$.  We also assume that $\cU_0$ smooth
and that it is supported in the interval $[0,1]$.

Summing up, the equations (\ref{u}), (\ref{pol_lin}) together with the
boundary condition (\ref{bound_cond_lin}) and the initial conditions
(\ref{init_lin}) form the initial-boundary value problem (IBVP)
describing the short-pulse propagation.
\begin{remark}
In general one can also expect an instantaneous contribution
to the polarization -- we ignore this case here since it can
be incorporated simply by changing the coefficient
in front of the term $u_{tt}$.
\end{remark}

We prove two approximation results about the solutions of
the IBVP \reff{u}.  Recall that from the formal derivation of
the short pulse equation in \cite{schaefer-wayne:2003}, we expect
that approximation of solutions of \reff{u} by the (linearized)
short pulse equation should be valid for times $\cO(\frac{1}{\epsilon})$.
In fact, following the usual convention in nonlinear optics
the multiple scale expansion in \cite{schaefer-wayne:2003} was
made in terms of a long {\em space} scale, $\epsilon x$,
rather than a long time scale $\epsilon t$, and thus one
might expect an approximation result valid over {\em length}
scales of $\cO(1/\epsilon)$.  However, since pulses in 
Eqn. \reff{u} travel with a speed $\cO(1)$, we can translate
our approximation result into a result valid over long time scales.
Our first result shows that for times of this order
we can approximate solutions of \reff{u} by solutions 
of the single equation
\begin{equation}\label{ux}
\tu_{xx} =  \tu_{tt} + \chi_0 \tu\ ,
\end{equation}
with initial conditions $\tu(x,0)=\tu_t(x,0)=0$ and boundary
condition $\tu(0,t) = U_0(t)$.  Note that this corresponds precisely
to the approximation in equation (5) in \cite{schaefer-wayne:2003}
and thus gives a rigorous justification of the heuristic
argument of that paper.
More precisely we prove the following:
\begin{proposition}\label{approx1}  Let $T_0 > 0$ be fixed.  Then there
exist $\epsilon_0>0$ and $C_0>0$ such that if $0 < \epsilon
< \epsilon_0$, and  $u(x,t)$ satisfies the IBVP 
\{(\ref{u}), (\ref{pol_lin}), (\ref{bound_cond_lin}), (\ref{init_lin})\}
 and $\tu(x,t)$ satisfies \reff{ux}, with zero initial data
and the boundary condition (\ref{bound_cond_lin}) then
\begin{equation} 
\sup_{0 \le t \le T_0/\epsilon}(\sup_{x > 0}|u(x,t)-\tu(x,t)|)
\le C_0 \epsilon^{1/2}\ .
\end{equation}
\end{proposition}

In this context we can also show that the linearized
version of the  short pulse
equation derived in \cite{schaefer-wayne:2003} correctly describes the 
propagation of solutions in either \reff{u} or \reff{ux}.
The linearized short pulse equation describing the 
evolution of a function of two variables $\cU=\cU(\phi,T)$ is
\begin{equation}\label{pulse}
2 \pp \partial_T \cU = \chi_0 \cU \ .
\end{equation}
We also prove
\begin{proposition}\label{approx2}  Let $T_0 > 0$ be fixed.  Then there
exist $\epsilon_0>0$ and $C_0>0$ such that if
$0 < \epsilon
< \epsilon_0$, and   $u(x,t)$ satisfies
 the IBVP 
\{(\ref{u}), (\ref{pol_lin}), (\ref{bound_cond_lin}), (\ref{init_lin})\}
then there exists a solution of the pulse equation
\reff{pulse}, $\cU(\phi,T)$ such that 
\begin{equation} \label{lin_ansatz_evolve}
\sup_{1 \le t \le T_0/\epsilon}(\sup_{x > 0}|u(x,t)-\cU(\tm,\T)|)
\le C_0 \epsilon^{1/2}\ .
\end{equation}
\end{proposition}
\begin{remark} Note that in this proposition we do not
begin to compare the solution of the ``true'' evolution
\reff{u} with the solution of the pulse equation until
a time $t > 1$ -- i.e. until the pulse has been
injected at the left boundary of the domain.  This is
because we don't expect the pulse equation to describe
the evolution before there is a pulse present
in the system!  
\end{remark}
\begin{remark} In both of these propositions the constants $C_0$
depend on the profile of the injected pulse, $\cU_0$, in
a way we make precise in the proof.
\end{remark}
We note that in \cite{alterman-rauch:2000b}, Alterman and Rauch
study in detail the properties of a linear short pulse equation
similar to \reff{pulse}, but lacking the term $\chi_0 \cU$.

\section{Proofs of the approximation results}
We begin with the proof of Proposition \ref{approx1}.
In the proofs of both propositions we will work with the
Fourier sine and cosine transforms.  Given a function
$v(x)$ defined on the positive half-line we define:
\begin{eqnarray}\label{FTdef}
&& \fs{v}(k) =  \int_0^{\infty} \sin(kx) v(x) dx\ , \\
&& \fc{v}(k) =  \int_0^{\infty} \cos(kx) v(x) dx\ .
\end{eqnarray}
We will also use the Laplace transform which we denote by
$\cL$.
If we take both the Fourier-sine and Laplace transforms
of \reff{u} and use the boundary and initial conditions we
find
\begin{eqnarray}\label{ulf}
&& -k^2 \cL[\fs{u}] + k \cL[U_0] =
s^2 \cL[\fs{u}] + s^2 \cL[\fs{p}]\ , \\
&& (s^2 + s\Gamma + \omega_0^2) \cL[\fs{p}] =
\chi_0 \cL[\fs{u}]\ .
\end{eqnarray}
We can combine these two expressions to obtain a single
equation for $\cL[\fs{u}]$, namely
\begin{equation}\label{flu}
\cL[\fs{u}](k,s) = \left\{
\frac{k}{(k^2+s^2) + \left(\frac{\chi_0 s^2}{
s^2 + s\Gamma + \omega_0^2}  \right) } \right\}
\cL[U_0](s)\ .
\end{equation}

If we now take the inverse Laplace transform of this expression
and use the fact that $U_0(t) = \cU(\frac{t}{\epsilon})$ we find that
\begin{equation}\label{ufs}
\fs{u}(k,t) = \epsilon \intt
\cL^{-1} \left\{
\frac{k}{(k^2+s^2) + \left(\frac{\chi_0 s^2}{
s^2 + s\Gamma + \omega_0^2}  \right) } \right\} (\epsilon[
\frac{t}{\epsilon}- \sigma])\cU_0(\sigma) d \sigma\ .
\end{equation}
We now rewrite this expression with the aid of the following
standard lemma about Laplace transforms:
\begin{lemma}\label{laplace} If $F(s) = \cL[f](s)$, then
$$
\epsilon f(\epsilon t) = \cL^{-1} [F(\frac{\cdot}{\epsilon})](t)\ .
$$
\end{lemma}
Using Lemma \ref{laplace} and defining $p = \epsilon k$, we
find that 
\begin{equation}\label{fsrescale}
\fs{u}(\frac{p}{\epsilon},t)=  \intt
\cL^{-1}[F(s,p;\epsilon)](\frac{t}{\epsilon}- \sigma)
\cU_0(\sigma) d \sigma\ ,
\end{equation}
where 
\begin{equation}\label{Fdef}
F(s,p;\epsilon) =
\frac{(\epsilon p)(s^2+\epsilon s \Gamma + \epsilon^2 \omega_0^2)}{
(s^2+\epsilon s \Gamma + \epsilon^2 \omega_0^2)(p^2+s^2)+
\epsilon^2 \chi_0s^2}\ .
\end{equation}
One can compute the inverse Laplace transform of $F$ via
it's partial fraction expansion and for that we need to
find the roots of the polynomial in the denominator of $F$, i.e.
\begin{equation}\label{Qdef}
Q(s;p,\epsilon) = (s^2+\epsilon s \Gamma + \epsilon^2 \omega_0^2)(p^2+s^2)+
\epsilon^2 \chi_0 s^2\ .
\end{equation}
To this end we use the following series of Lemmas whose
proofs are elementary but somewhat involved. Hence
we relegate the proofs to Appendix \ref{eigen}.
\begin{lemma}\label{neg} For all values of $p$ and $\epsilon$ all
the eigenvalues of $Q$ have non-positive real part.
\end{lemma}
\begin{lemma}\label{s0} There exist $\epsilon_0>0$ and $C_0, C_1 >0$ such
that for $|\epsilon| < \epsilon_0$ and $p > C_0 \epsilon$,
$Q$ has a pair of roots of the form
$$
s^0_{\pm} = -\frac{\epsilon}{2}(\Gamma \pm \sqrt{\Gamma^2
- 4 \omega_0^2}) + \epsilon \sigma^0_{\pm}(p)
$$
with
$$
|\sigma^0_{\pm}(p)| \le \frac{C_1 \epsilon^2}{p^2}\ .
$$
\end{lemma}
\begin{lemma}\label{s1} There exists $\epsilon_0>0$ and $C_0 >0$ such
that for $|\epsilon| < \epsilon_0$ and $p > C_0 {\epsilon}$,
$Q$ has a pair of roots of the form
$$
s^1_{\pm} = \pm i \sqrt{p^2 + \epsilon^2 \chi_0}
+\sigma^1_{\pm}(p)
$$
with
$$
|\sigma^1_{\pm}(p)| \le \frac{C_0 \epsilon^3}{p^2}\ .
$$
Furthermore, the real part of $\sigma^1_{\pm}$ is negative.
\end{lemma}
\begin{lemma}\label{smallp} Let $\epsilon_0$ and $C_0$ be as
in Lemma \ref{s1}. There exists $c_{min} >0$ such
that for $p < C_0 \epsilon$ the
roots of $Q$ can be written as
$s^0_{\pm}(p) = \epsilon \ts^0_{\pm} (p/\epsilon)$ and 
$s^1_{\pm}(p) = \epsilon \ts^1_{\pm} (p/\epsilon)$ with
$\ts^0_{\pm}(q)$ and $\ts^1_{\pm}(q)$ all distinct.
Furthermore,
$$
\min( |\ts^0_{+}(q)-\ts^0_{-}(q)|, |\ts^0_{\pm}(q)
- \ts^1_{\pm}(q)|) \ge c_{min}\ ,
$$
while
$$
|\ts^1_{+}(q)-\ts^1_{-}(q)| \ge c_{min} q\ .
$$
\end{lemma}
\begin{remark} Note that a corollary of the proof of 
Lemma \ref{smallp} is that for $p>0$ the roots of $Q$ are all
distinct.
\end{remark}

With these estimates on the eigenvalues of $Q$ in hand we
now construct the partial fraction expansion of $F$.  Note
that since the eigenvalues of $Q$ depend continuously
on $p$ (and $\epsilon$) we can label the eigenvalues
as $s^0_{\pm}(p)$ and $s^1_{\pm}(p)$ for all $p\ge 0$,
and then one can write
\begin{equation}\label{partial_frac}
F(s,p;\epsilon) =
\frac{A^0_+(p)}{s-s^0_+(p)}+\frac{A^0_-(p)}{s-s^0_-(p)}+
\frac{A^1_+(p)}{s-s^1_+(p)}+\frac{A^1_-(p)}{s-s^1_-(p)}\ .
\end{equation}

If we also write
\begin{equation}\label{factor}
F(s,p;\epsilon) =\frac{(\epsilon p) (s^2+\epsilon s \Gamma +
\epsilon^2 \chi_0)}{(s-s^0_+(p))(s-s^0_-(p))
(s-s^1_+(p))(s-s^1_-(p))}\ ,
\end{equation}
then we see that $A^0_+$ has the form
\begin{equation}\label{A0plus}
A_+^0 = \frac{(\epsilon p) ((s^0_+)^2+\epsilon (s^0_+) \Gamma +
\epsilon^2 \omega_0^2)}{((s^0_+)-s^0_-)
((s^0_+)-s^1_+)((s^0_+)-s^1_-)}\ .
\end{equation}
First note that if $p < C_0 \epsilon$, the numerator of this
expression can be bounded by $C \epsilon^4$ by Lemma \ref{smallp}
while the same lemma guarantees that the denominator is
bounded below by $c \epsilon^3$ for some $c>0$.  Thus
for $p$ in this range $|A^0_+| \le C\epsilon$.

Now suppose that $p > C_0 \epsilon$.
Since $s^0_+$ is a root of $Q$, we have
$$
|((s^0_+)^2+\epsilon (s^0_+) \Gamma + \epsilon^2 \omega_0^2)|
= \left|\frac{\epsilon^2 \chi_0 (s^0_+)^2 }{(s^0_+)^2+p^2}\right|.
$$
Now we see that this expression is bounded by
$C \epsilon^4/ p^2$.  Using the asymptotic expressions for the
roots of $Q$ coming from Lemma \ref{s0} and \ref{s1} 
the denominator
in \reff{A0plus} can be bounded from below by $\epsilon p^2$ in
this range and hence we can bound
$$
|A^0_+| \le \frac{C \epsilon^3}{p^2}\ ,
$$
for $p$ in this range.  We can combine the estimates on 
$|A^0_+|$ in these two ranges of $p$ along with identical
estimates on $A^0_-$ to obtain
\begin{lemma}\label{A0est} There exists a constant $C_A >0$ such that for
all $p>0$,
$$
|A^0_{\pm}(p)| \le \frac{ C_A \epsilon^3}{p^2+\epsilon^2}\ .
$$
\end{lemma}

We now estimate the coefficients $A^1_{\pm}$.  
Following an argument similar to that above we find
\begin{equation}\label{A1plus}
A^1_{+} = \frac{(\epsilon p) ((s^1_+)^2+\epsilon (s^1_+) \Gamma +
\epsilon^2 \omega_0^2)}{((s^1_+)-s^1_-)
((s^1_+)-s^0_+)((s^1_+)-s^0_-)}\ .
\end{equation}
For $p < C_0 \epsilon$, using the asymptotic values of the roots $s^1_+$ and $s^0_{\pm}$
given in Lemma \ref{smallp} we see that 
$$
|A^1_{+}(p)| \le C \epsilon\ .
$$  
For $p \ge C_0 \epsilon$ we use Lemma \ref{s0} and
\ref{s1} to rewrite
\begin{eqnarray}\label{A1est}\nonumber
&& \frac{(s^1_+)^2+\epsilon (s^1_+) \Gamma +
\epsilon^2 \omega_0^2}{((s^1_+)-s^0_+)((s^1_+)-s^0_-)} =
\frac{((s^1_+)^2+\epsilon (s^1_+) \Gamma +
\epsilon^2 \omega_0^2)}{((s^1_+)^2+ \epsilon \Gamma s^1_+  
- \epsilon (\sigma^0_{+} + \sigma^0_{-})s^1_+
+ s^0_{+} s^0_{-}
)} \\
&& \qquad = 1+ \cE(p;\epsilon)\ ,
\end{eqnarray}
where
\begin{equation}\label{Eest}
|\cE(p;\epsilon)| \le \frac{C_E \epsilon^2}{p^2 +\epsilon^2}\ .
\end{equation}
Thus,
\begin{equation}
A^1_{+} = \frac{\epsilon p (1+\cE(p;\epsilon))}{(2i\sqrt{p^2+
\epsilon^2 \chi_0} + (\sigma^1_{+}-\sigma^1_{-}))}\ ,
\end{equation}
or if we write
\begin{equation}
A^1_{+} = \frac{\epsilon p}{2i\sqrt{p^2+
\epsilon^2 \chi_0}} + \Delta A^1_{+}\ ,
\end{equation}
we see that for $p > C_0 \epsilon$, (and $C_0$ sufficiently large)
\begin{equation}
|\Delta A^1_{+}(p)| \le \frac{ C_A \epsilon^3 p}{(p^2 + \epsilon^2)^{3/2}}\ .
\end{equation}
Similar estimates hold for $A^1_{-}$ and we have established:
\begin{lemma}\label{Alcoef} There exists a constant $C_A>0$ such
that for $p < C_0 \epsilon$,
$$
|A^1_{\pm}(p)| \le C_A \epsilon\ ,
$$
while for $p > C_0 \epsilon$, 
$$
A^1_{\pm}(p) = \frac{\pm \epsilon p}{2i\sqrt{p^2+
\epsilon^2 \chi_0}} + \Delta A^1_{\pm}\ ,
$$
with
$$
|\Delta A^1_{\pm}(p)| \le \frac{ C_A \epsilon^3 p}{(p^2 + \epsilon^2)^{3/2}}\ .
$$
\end{lemma}

With these estimates on the coefficients in the partial
fraction decomposition of $F(s,p;\epsilon)$ in hand, we now return
to the task of computing the solution $u(x,t)$ of (\ref{u}).
From \reff{fsrescale} and \reff{partial_frac} we see that
\begin{eqnarray}\label{uform} \nonumber
&& u(x,t) = \frac{2}{\pi \epsilon} \int_0^{\infty} \sin(p (\frac{x}{\epsilon}))
\intt A^0_{+}(p) e^{s^0_{+}(p) (\argt)} \cU_0(\sigma) d\sigma dp \\ \nonumber
&&\qquad +\frac{2}{\pi \epsilon} \int_0^{\infty} \sin(p (\frac{x}{\epsilon}))
\intt A^0_{-}(p) e^{s^0_{-}(p) (\argt)} \cU_0(\sigma) d\sigma dp \\
&&\qquad + \frac{2}{\pi \epsilon} \int_0^{\infty} \sin(p (\frac{x}{\epsilon}))
\intt A^1_{+}(p) e^{s^1_{+}(p) (\argt)} \cU_0(\sigma) d\sigma dp \\ \nonumber
&&\qquad + \frac{2}{\pi \epsilon} \int_0^{\infty} \sin(p (\frac{x}{\epsilon}))
\intt A^1_{-}(p) e^{s^1_{-}(p) (\argt)} \cU_0(\sigma) d\sigma dp \ .
\end{eqnarray}

We can immediately bound the first two terms on the
right hand side of \reff{uform} by using the facts
that the real parts of $s^0_{\pm}$ are negative, so that
the exponential factor is bounded by $1$, as is the factor
of $\sin(px/\epsilon)$, and bounding $A^0_{\pm}$ by
the bound in Lemma \ref{A0est}.  Integrating over $p$ and $\sigma$
we then see that these two lines are bounded by
$C \epsilon \| \cU_0 \|_{L^1}$.  We now turn to the last two
lines in \reff{uform}. First of all, rewriting them
with the aid of Lemma \ref{s1} and \ref{Alcoef} as
\begin{eqnarray}\nonumber
&& \frac{2}{\pi \epsilon} \int_{p=C_1 \sqrt{\epsilon}}^{\infty}
 \intt \frac{\epsilon p}{\sqrt{
p^2 + \epsilon^2 \chi_0}} 
\sin(p (\frac{x}{\epsilon})) \sin(\sqrt{p^2 + \epsilon^2 \chi_0}
(\argt)) \cU_0(\sigma) d\sigma dp \\ \nonumber  &&
+ \frac{2}{\pi \epsilon} \int_{p= C_1 \sqrt{\epsilon}}^{\infty}
\intt (\Delta A^1_{+}(p)
e^{s^1_{+}(p) (\argt)} \\ \nonumber &&\qquad \qquad  + \Delta A^1_{-}(p)
e^{s^1_{-}(p) (\argt)}) \sin(p(\frac{x}{\epsilon})) \cU_0(\sigma)
d\sigma dp \\ \nonumber &&
+  \frac{2}{\pi \epsilon} \int_{p=C_1 \sqrt{\epsilon}}^{\infty} \intt 
\sin(p (\frac{x}{\epsilon})) \big\{\frac{1}{2i} \frac{\epsilon p}{
\sqrt{p^2+\epsilon^2 \chi_0}} e^{i \sqrt{p^2+\epsilon^2 \chi_0}}
(e^{\sigma_+^1(\frac{t}{\epsilon}-\sigma)}-1) \\  &&
\qquad \qquad  -
\frac{1}{2i} \frac{\epsilon p}{
\sqrt{p^2+\epsilon^2 \chi_0}} e^{- i \sqrt{p^2+\epsilon^2 \chi_0}}
(e^{\sigma_{-}^1(\frac{t}{\epsilon}-\sigma)}-1) \big\}\cU_0(\sigma)
d\sigma dp \\ \nonumber && +  \frac{2}{\pi \epsilon} 
\int_0^{p=C_1 \sqrt{\epsilon}} \intt
\sin(p (\frac{x}{\epsilon})) \left\{  A^1_{+}(p)
e^{s^1_{+}(p) (\argt)} +  A^1_{-}(p)
e^{s^1_{-}(p) (\argt)} \right\} \cU_0(\sigma)
d\sigma dp .
\end{eqnarray}
The last of these integrals can be immediately bounded by
$C_A \sqrt{\epsilon} \| \cU_0 \|_{L^1}$.  The next to last
integral is estimated by using Lemma \ref{s1} and the
fact that for $0 \le t \le T_0/\epsilon$,
$$
|e^{\sigma^1_{\pm}(\argt)} - 1| \le \frac{C \epsilon}{p^2}\ .
$$
With this estimate the integral can be bounded by
$$
C \int_{C_1\sqrt{\epsilon}}^{\infty} 
\left(\frac{p}{\sqrt{p^2 + \epsilon^2 \chi_0}}
+ \frac{\epsilon^3 p}{(p^2 + \epsilon^2)^{3/2}}\right)
\frac{\epsilon}{p^2} \|\cU_0 \|_{L^1} dp
\le C \sqrt{\epsilon} \| \cU_0 \|_{L^1}\ .
$$
Finally, in the second term we use the bounds on $\Delta
A_{\pm}^1$ from Lemma \ref{Alcoef} to estimate this integral
by
$$
C \epsilon^2 \int_{C_1\sqrt{\epsilon}}^{\infty}
\frac{p}{(p^2 + \epsilon^2)^{3/2}} dp \| \cU_0 \|_{L^1}
\le C \epsilon \| \cU_0 \|_{L^1}\ .
$$
Combining these estimates with those on the first
two integrals in \reff{uform} we see that 
\begin{eqnarray}\label{inter1}
&& \sup_{0 \le t \le T_0/\epsilon}\left(\sup_{x > 0}\left|u(x,t)
- \frac{2}{\pi \epsilon} \int_{p=C_1 \sqrt{\epsilon}}^{\infty}
 \intt \frac{\epsilon p}{\sqrt{
p^2 + \epsilon^2 \chi_0}} 
\sin(p (\frac{x}{\epsilon}))\times \right.\right.\\ \nonumber
&& \qquad \qquad  \times \left.\left.\sin\left(\sqrt{p^2 + \epsilon^2 \chi_0}
(\argt)\right) \cU_0(\sigma) d\sigma dp \right|\right) \le C \sqrt{\epsilon} \| \cU_0 \|_{L^1}\ .
\end{eqnarray}
We now note that since 
\begin{eqnarray}\nonumber
&& \left|\frac{2}{\pi \epsilon} \int_{p=0}^{C_1 \sqrt{\epsilon}}
 \intt \frac{\epsilon p}{\sqrt{
p^2 + \epsilon^2 \chi_0}} 
\sin(p (\frac{x}{\epsilon})) \sin(\sqrt{p^2 + \epsilon^2 \chi_0}
(\argt)) \cU_0(\sigma) d\sigma dp \right|\\ 
&& \qquad \qquad  \le C \sqrt{\epsilon} 
 \| \cU_0 \|_{L^1} 
\end{eqnarray}
we can subtract it from the left hand side of \reff{inter1}
without changing the bound on the right hand side of this
expression.  That is, we can bound the difference between
$u(x,t)$ and the integral
$$
\frac{2}{\pi \epsilon}\int_{p=0}^{\infty}
 \intt \frac{\epsilon p}{\sqrt{
p^2 + \epsilon^2 \chi_0}} 
\sin(p (\frac{x}{\epsilon})) \sin(\sqrt{p^2 + \epsilon^2 \chi_0}
(\argt)) \cU_0(\sigma) d\sigma dp.
$$
by $C \sqrt{\epsilon} \| \cU_0 \|_L^1$.

By taking the Fourier sine transform of \reff{ux} we see
that this integral is exactly $\tu(x,t)$, the solution
of \reff{ux} and thus we have completed the proof 
of Proposition \ref{approx1}.
\qed

We next prove Proposition \ref{approx2}.  By the results established
so far it suffices to prove that the solution $\tu(x,t)$ of 
\reff{ux} can be approximated by a solution of the pulse equation
\reff{pulse} over the relevant time intervals.
First note that using trigonometric identities for the
sine and cosine we can rewrite
\begin{equation}\label{LR}
\tu(x,t) = u^L(x,t)+u^R(x,t) \,
\end{equation}
where 
\begin{equation}\label{uR}
u^R(x,t) = \frac{1}{\pi \epsilon} \int_0^{\infty}
\intt \frac{\epsilon p}{\ome(p)} \cos(\frac{1}{\epsilon}(p x
- \ome(p) t) + \sigma \ome(p)) \cU_0(\sigma) d\sigma dp\ ,
\end{equation}
and 
\begin{equation}\label{uL}
u^L(x,t) = -\frac{1}{\pi \epsilon} \int_0^{\infty}
\intt \frac{\epsilon p}{\ome(p)} \cos(\frac{1}{\epsilon}(p x
+ \ome(p) t)- \sigma \ome(p)) \cU_0(\sigma) d\sigma dp\ ,
\end{equation}
where $\ome(p) = \sqrt{p^2 + \epsilon^2 \chi_0}$.

Roughly speaking, $u^R$ and $u^L$ represent the left and
right moving parts of the pulse.  In particular, for
$t >1$, we expect that the left moving part of the solution
will no longer be relevant since we are only interested in 
the solution for $x > 0$.  To prove this we use
another trigonometric identity to rewrite $u^L$
as
\begin{eqnarray}\label{uL2} \nonumber
&&u^L(x,t) = -\frac{1}{\pi} \int_0^{\infty}
\frac{p}{\ome(p)} \cos(\frac{1}{\epsilon}(px +
\ome(p) t) \hat{\cU}_0^c(\ome(p)) dp \\ && 
\qquad \quad  - \frac{1}{\pi} \int_0^{\infty}
\frac{p}{\ome(p)} \sin(\frac{1}{\epsilon}(px +
\ome(p) t) \hat{\cU}_0^s(\ome(p)) dp\ ,
\end{eqnarray}
and we recall that $\hat{\cU}_0^c$ and $\hat{\cU}_0^s$
are the cosine and sine transforms of the boundary
data $\cU_0$.  (We have used here the fact that since
the limit on the $\sigma$ integral exceeds the limits
on the support of $\cU_0$ we can integrate from 
$0$ to $\infty$.)
We now prove that both of these terms are $\cO(\sqrt{\epsilon})$ in
the $L^{\infty}$ norm and thus can be ignored to the
order of approximation that we are concerned with.

We'll consider in detail the first of the two terms in \reff{uL2}.
The second is handled in an almost identical fashion and we leave
the details as an exercise.  Rewrite that integral using
a trigonometric identity as
\begin{eqnarray}\label{uL3}\nonumber
&& -\frac{1}{\pi} \int_0^{\infty} \{ \cos(\frac{p}{\epsilon}
(x+t) ) \cos(\frac{t}{\epsilon}(\omega_{\epsilon}(p)-p) \}
\frac{p}{\omega_{\epsilon}(p)} \hat{\cU}_0^c(\omega_{\epsilon}(p)) dp \\
&&\frac{1}{\pi} \int_0^{\infty} \{ \sin(\frac{p}{\epsilon}
(x+t) ) \sin(\frac{t}{\epsilon}(\omega_{\epsilon}(p)-p) \}
\frac{p}{\omega_{\epsilon}(p)} \hat{\cU}_0^c(\omega_{\epsilon}(p)) dp\ .
\end{eqnarray}
Once again these two integrals are estimated in an almost identical
fashion so we provide the details for the first and leave the 
second as an exercise.  Integrating by parts, the first integral
becomes
\begin{eqnarray}\label{uL4}\nonumber
&& \frac{1}{\pi} \int_0^{\infty} \frac{\epsilon}{x+t} 
\sin(\frac{p}{\epsilon}
(x+t) ) \big\{ \frac{t}{\epsilon}(\omega^{\prime}_{\epsilon}(p)-1)
\frac{p}{\omega_{\epsilon}(p)} \sin(\frac{t}{\epsilon}(\omega_{\epsilon}(p)-p))
 \hat{\cU}_0^c(\omega_{\epsilon}(p)) \\
&& \qquad
-\cos(\frac{t}{\epsilon}(\omega_{\epsilon}(p)-p))[\frac{
\omega_{\epsilon}(p) - p \omega^{\prime}_{\epsilon}(p)}{
(\omega_{\epsilon}(p))^2}]
\hat{\cU}_0^c(\omega_{\epsilon}(p) \\ \nonumber
&& \qquad \qquad  -\cos(\frac{t}{\epsilon}(\omega_{\epsilon}(p)-p))
\frac{p}{\omega_{\epsilon}(p)} \omega_{\epsilon}^{\prime}(p)
\hat{\cU}_0^{c \prime}(\omega_{\epsilon}(p)) \big\} dp\ .
\end{eqnarray}
Note that there exists a constant
$C_1>0$, independent of $\epsilon$ such that the
the various quotients appearing in the integrand
of \reff{uL4} can be bounded as follows:
\begin{displaymath}
p\big| \frac{\omega^{\prime}_{\epsilon}(p)-1}{\omega_{\epsilon}(p)}\big|
\le \begin{cases} C_1 &\text{ for all }  p>0, \\
\frac{C_1 \epsilon^2}{p^2} 
&\text{ for all }  p > C_0 \sqrt{\epsilon}. \end{cases}
\end{displaymath}
\begin{displaymath}
\big| \frac{\omega_{\epsilon}(p) - p\omega^{\prime}_{\epsilon}(p)}{
(\omega_{\epsilon}(p))^2}\big|
\le \begin{cases} \frac{C_1}{\epsilon} &\text{ for all }  p>0, \\
\frac{C_1 \epsilon^2}{p^3} 
&\text{ for all }  p > C_0 \sqrt{\epsilon}. \end{cases}
\end{displaymath}
\begin{equation*}
\big| \frac{p \omega^{\prime}_{\epsilon}(p)}{\omega_{\epsilon}(p)} \big|
\le C_1~~{\rm for~all~}p>0
\end{equation*}

Thus, bounding the factors of sine and cosine in \reff{uL4}
by $1$ we see that this integral can be bounded by
\begin{eqnarray}\label{uL5}
&& \frac{C}{x+t} \int_0^{C_0 \sqrt{\epsilon}} (|t|+1) 
|\hat{\cU}_0^{c}(\omega_{\epsilon}(p))| dp \\ \nonumber
&& \quad
+ \frac{C\epsilon^2 }{x+t} \int_{C_0 \sqrt{\epsilon}}^{\infty}
(|t|+1) (\frac{1}{p^2}+ \frac{1}{p^3})
|\hat{\cU}_0^{c}(\omega_{\epsilon}(p))| dp
+ \frac{C \epsilon}{x+t} \int_0^{\infty} 
|\hat{\cU}_0^{c\prime }(\omega_{\epsilon}(p))| dp .
\end{eqnarray}

In the first two of these integrals we bound
$|\hat{\cU}_0^{c}(\omega_{\epsilon}(p))|$ by $C \|\cU_0 \|_{L^1}$
Thus, these first two integrals can be bounded by
$C \sqrt{\epsilon} \| \cU_0 \|_{L^1}$.  To bound the
final integral write it as the sum
$$\int_0^{1} 
|\hat{\cU}_0^{c\prime }(\omega_{\epsilon}(p))| dp
+ \int_1^{\infty} 
|\hat{\cU}_0^{c\prime }(\omega_{\epsilon}(p))| dp
$$
The first integral can again be bounded by
$C \| \cU_0 \|_{L^1}$, while the second is bounded by
$C \int_1^{\infty} 
|\hat{\cU}_0^{c\prime }(\xi)| d\xi$, by making the change
of variables $\xi =\omega_{\epsilon}(p)$.  This integral can
be bounded by
$C (\int_1^{\infty} (1 + \xi^2) | \hat{\cU}_0^{c\prime }(\xi)|^2 d\xi)^{1/2}$
by the Cauchy-Schwartz.  Applying Parseval's equality and
the fact that $\cU_0$ has finite support this integral is bounded
by $\| \cU_0\|_{H^1}$.  Note that since $\cU_0$ has finite support,
one can also bound the $L^1$ norm of $\cU_0$ by a constant times
the $H^1$ norm, and thus,
\reff{uL5} is bounded by $\frac{C(t+1)}{x+t} \sqrt{\epsilon}
\|\cU_0 \|_{H^1}$  A similar estimate applies to the remaining
terms in $u^L$ and so we conclude that
\begin{equation}\label{uLdiff}
\sup_{t \ge 1} \sup_{x > 0} |\tilde{u} - u^R(x,t)|
\le C \sqrt{\epsilon} \|\cU_0 \|_{H^1}\ .
\end{equation}

We now examine $u^R$ more closely and show that it can
be approximated by a solution of the pulse equation.
Begin by writing it as
\begin{eqnarray}\label{uR2} \nonumber
&&u^R(x,t) = \frac{1}{\pi} \int_0^{\infty}
\frac{p}{\ome(p)} \cos(\frac{1}{\epsilon}(px -
\ome(p) t) \hat{\cU}_0^c(\ome(p)) dp \\ && 
\qquad \quad +\frac{1}{\pi} \int_0^{\infty}
\frac{p}{\ome(p)} \sin(\frac{1}{\epsilon}(px 
- \ome(p) t) \hat{\cU}_0^s(\ome(p)) dp\ .
\end{eqnarray}
Now define $\phi=\left(\frac{x-t}{\epsilon}\right)$ and $T=\epsilon t$.
Then
\begin{eqnarray}\label{uRnew} \nonumber
&& \cU^R(\phi,T) \equiv u^R(x(\phi,T),t(\phi,T)) = \\ && \qquad
 \frac{1}{\pi} \int_0^{\infty}
\frac{p}{\ome(p)} \cos(p\phi +(p-\ome(p))\frac{T}{\epsilon^2}) 
\hat{\cU}_0^c(\ome(p)) dp \\ \nonumber
&& \qquad \quad +\frac{1}{\pi} \int_0^{\infty}
\frac{p}{\ome(p)} \sin(p\phi +(p-\ome(p))\frac{T}{\epsilon^2}) 
\hat{\cU}_0^s(\ome(p)) dp\ ,
\end{eqnarray}
We now define
\begin{eqnarray}\label{Udef}\nonumber
&& \cU(\phi,T)=
 \frac{1}{\pi} \int_{\epsilon^{1/2}}^{\infty}
\frac{p}{\ome(p)} \cos(p\phi - \frac{\chi_0 T}{2p}) 
\hat{\cU}_0^c(\ome(p)) dp \\ 
&& \qquad \quad +\frac{1}{\pi} \int_{\epsilon^{1/2}}^{\infty}
\frac{p}{\ome(p)} \sin(p\phi  - \frac{\chi_0 T}{2p})
\hat{\cU}_0^s(\ome(p)) dp\ .
\end{eqnarray}
Note that by an easy and explicit computation
$\cU(\phi,T)$ satisfies the linearized pulse equation
\reff{pulse}, hence, Proposition \ref{approx2} will follow
if we can show that $\Delta \cU = \cU^R - \cU$ is small.
Subtracting \reff{Udef} from \reff{uRnew} we obtain
\begin{eqnarray}\label{DUdef}
\nonumber
&&\Delta\cU(\phi,T)=
 \frac{1}{\pi} \int_{\epsilon^{1/2}}^{\infty}
\frac{p}{\ome(p)}\left\{ \cos(p\phi +(p-\ome(p))\frac{T}{\epsilon^2})
- \cos(p\phi - \frac{\chi_0 T}{2p}) \right\}
\hat{\cU}_0^c(\ome(p)) dp \\ \nonumber
&& \quad+\frac{1}{\pi} \int_{\epsilon^{1/2}}^{\infty}
\frac{p}{\ome(p)}\left\{  \sin(p\phi +(p-\ome(p))\frac{T}{\epsilon^2})
- \sin(p\phi  - \frac{\chi_0 T}{2p})\right\}
\hat{\cU}_0^s(\ome(p)) dp \\ \nonumber
&& \qquad \quad
+\frac{1}{\pi} \int_0^{\epsilon^{1/2}}
\frac{p}{\ome(p)} \cos(p\phi +(p-\ome(p))\frac{T}{\epsilon^2}) 
\hat{\cU}_0^c(\ome(p)) dp \\
&& \qquad \quad +\frac{1}{\pi} \int_0^{\epsilon^{1/2}}
\frac{p}{\ome(p)} \sin(p\phi +(p-\ome(p))\frac{T}{\epsilon^2}) 
\hat{\cU}_0^s(\ome(p)) dp\ .
\end{eqnarray}
Note that the integrals over $[0,\sqrt{\epsilon}]$ can be
easily bounded by noting that $|p/\ome(p)| \le 1$, hence both
of these integrals are bounded by 
$C \sqrt{\epsilon} (|\hat{\cU}_0^s|_{L^{\infty}} 
+|\hat{\cU}_0^c|_{L^{\infty}})$.
The two remaining integrals are bounded in a similar fashion -- we give the details of the  bound
on the integral containing
the difference of cosines and leave the other as an exercise. 
By the mean value theorem there exists some $\xi$ such
that
\begin{eqnarray}\label{taylor}\nonumber
&& | \cos(p\phi +(p-\ome(p))\frac{T}{\epsilon^2})
- \cos(p\phi - \frac{\chi_0 T}{2p})| =
|\sin(\xi) \left\{ (p-\ome(p))\frac{T}{\epsilon^2}
+ \frac{\chi_0 T}{2 p}\right\}| \\\nonumber
&& \qquad \quad \le C T \frac{\chi_0^2 \epsilon^2}{p^3}\ ,
\end{eqnarray}
where the last inequality used Taylor's theorem to bound
$\left\{ (p-\ome(p))\frac{T}{\epsilon^2}
+ \frac{\chi_0 T}{2 p}\right\}$ and the fact that for
$p \ge \epsilon^{1/2}$, $\frac{\epsilon^3}{p^3} << 1$.
Inserting this estimate into the first of the integral
terms in \reff{DUdef} we see that it is bounded
by
$$
C T \int_{\epsilon^{1/2}}^{\infty}\frac{\chi_0^2 \epsilon^2}{p^2 \ome(p)}
\hat{\cU}_0^c(\ome(p)) dp \le C T \sqrt{\epsilon} |\hat{\cU}_0^c|_{L^{\infty}}
\ . 
$$
A similar estimate holds for the remaining term in the definition
of $\Delta\cU$ and Proposition \ref{approx2} follows.
\qed

\section{Approximating the nonlinear pulse dynamics using renormalization groups}

Summarizing the results of the previous section we now know
(rigorously) that if we ignore nonlinear effects we can approximate
the motion of a short pulse injected into one end of an optical fiber
by the linearized short-pulse equation (\ref{pulse}). Since nonlinear
effects are important for many optical phenomena, the next natural
step is to investigate how incorporating nonlinear terms into the
polarization affects (\ref{pulse}). Here, we consider as a first step
the simplest form of a nonlinear contribution $p_{\mathrm{nl}}$ given
by
\begin{equation}
p_{\mathrm{nl}}= \chi_3 u^3.
\end{equation}
In order to answer this question we start from a wave
equation similar to (\ref{ux}) with an additional nonlinear term
\begin{equation} \label{nl_max}
u_{xx} = u_{tt} + \chi_0 u + \chi_3 (u^3)_{tt}.
\end{equation}
\begin{remark}
The equation (\ref{nl_max}) can be derived from Maxwell's wave equation
\begin{displaymath}
u_{xx}-u_{tt}=(p_{\mathrm{lin}})_{tt}+(p_{\mathrm{nl}})_{tt}
\end{displaymath}
writing the linear part of the polarization as
\begin{displaymath}
p_{\mathrm{lin}}(x,t)= \int \chi(t-\tau)u(x,\tau)d\tau
\end{displaymath}
and making the approximation
\begin{displaymath}
\hat{\chi}(\omega)= -\frac{\chi_0}{\omega^2-i\Gamma\omega-\omega_0^2}\approx - \frac{\chi_0}{\omega^2}.
\end{displaymath}
This means physically that the frequency range of the pulse under
consideration is not only far from the resonance frequency of the material but also much larger than $\omega_0$. It is also possible to consider other forms of the susceptibility
in frequency domain, leading to different types of equations for the
short pulse. If we ignore the nonlinear term, then
Proposition \ref{approx1} shows that this approximation leads to a 
small error. The choice of nonlinear part of the susceptibility
corresponds to 
\begin{displaymath}
p_{\mathrm{nl}}(x,t)= \int \chi^{(3)}(t-\tau_1,t-\tau_2,t-\tau_3)u(x,\tau_1)u(x,\tau_2)u(x,\tau_3) d\tau_1 d\tau_2 d\tau_3
\end{displaymath}
with the assumption that the nonlinear contribution is instantaneous, hence
\begin{displaymath}
\chi^{(3)}(t-\tau_1,t-\tau_2,t-\tau_3) = \chi_3 \delta(t-\tau_1)\delta(t-\tau_2)\delta(t-\tau_3).
\end{displaymath}
On the basis of formal asymptotic calculations we believe that this is
the most important contribution to the nonlinearity and thus in the
present paper we limit ourselves to the consideration of this
case. However, we also stress that, for ultra-short pulses, it is very
interesting to extend the analysis to more complicated forms of the
nonlinear susceptibility \cite{boyd:1992} including delay in the
response of the material.
\end{remark}
As mentioned in the introduction, the standard model describing the
nonlinear pulse evolution is the cubic nonlinear Schr{\"o}dinger
equation (NLSE). To emphasize the different regimes to which the NLSE
and short pulse equations apply, we briefly review how one derives the
NLSE from (\ref{nl_max}). The main idea is that we assume a broad,
rather than a short, pulse in the sense that we introduce time scales
that are {\em slower} than the oscillations of the carrier wave that
is oscillating at a fixed frequency $\tilde
\omega$ with a wavenumber $\tilde \beta$. Therefore, the NLSE is an
equation describing the slowly varying amplitude of the optical
signal.

The separation of those time scales can be done by a usual expansion
in multiple scales \cite{newell-moloney:1992}. In the present work,
however, we utilize the so-called renormalization group (RG) method to
derive the NLSE. This perturbative technique was first developed by
Chen, Goldenfeld, and Oono as a tool for asymptotic analysis (see
\cite{chen-goldenfeld-etal:1994} and
\cite{chen-goldenfeld-etal:1996}). In
\cite{chen-goldenfeld-etal:1996}, the validity of the RG method has 
been justified by applying it to various examples of ordinary
differential equations involving multiple scales, boundary layers and
WKB analysis. See also, 
\cite{bricmont:1994}, \cite{bona:1996}, and \cite{deville-harkin-josic-kaper:preprint} for some examples of the
rigorous use of the renormalization group in the study of partial
differential equations. The mathematical study of this method has also
been presented by Ziane in \cite{ziane:2000}. The author explicitly
described the RG method in the general setting of autonomous nonlinear
systems of differential equations.

We will follow the approach given in \cite{ziane:2000} and explain how
to obtain the perturbative solution of Maxwell's equation. Although
the RG method will lead to the same results as in multiple scale
technique, it is worth mentioning that there are advantages of using
this method. First, the RG method does not require one to introduce all
the different scales in the beginning of the ansatz since these will
appear naturally in the {\it renormalization group equation}. This
implies that one can assume a naive perturbation series in any given
problems involving multiple scales. A second argument in favor of this
method is that the algebraic calculations are simpler than when other
perturbation techniques are used, especially when one considers higher
order approximations.

To see how the RG method works in the present case, we start from a
slightly more general form than (\ref{nl_max})
\begin{equation} \label{master}
u_{xx} = u_{tt} + \partial_{tt}\int \chi(t-\tau) u(x,\tau)d\tau + \chi_3 (u^3)_{tt}.
\end{equation}
We assume that the solution of (\ref{master}) is of small amplitude
and concentrated around the carrier frequency.
Because of the
oscillations of the carrier wave, in the Fourier domain the signal will
be concentrated around the frequencies $\tilde\oo$ and
$-\tilde\oo$. Therefore, we can write our solution as a wave packet
in the form of
\begin{equation}
u(x,t) = U(x,t)e^{i(\tilde{\bb}x-\tilde{\oo} t)} +
U^*(x,t)e^{-i(\tilde{\bb}x-\tilde{\oo}t)}. \label{wave_noscales}
\end{equation}
Taking the Fourier transform of (\ref{master}), we find
\begin{equation} \label{master_fourier}
\left(\frac{\ppy^2}{\ppy x^2} + 
\bb^2(\oo)\right)\hat{u}(x,\oo) = -\oo^2 \chi_3 \widehat{u^3}(x,\oo),
\end{equation}
where the wavenumber $\beta$ is given by 
\begin{equation} \label{disprelation}
\beta(\omega)=\omega\sqrt{1+\hat{\chi}(\omega)}.
\end{equation}
The main idea is now to make a
Taylor expansion of the dispersion $\beta$ around the carrier
frequency. This assumes that the signal is localized in Fourier
domain, corresponding to a slowly varying amplitude approximation in
time domain. Because of this {\em{local}} character of this expansion,
the specific form of $\chi(\omega)$ is not essential. The Taylor expansion of $\beta$ at $\tilde\omega$ yields
\begin{equation}
\beta^2(\omega) = \beta^2(\tilde\omega)+\left.\frac{\partial \beta^2}{\partial \omega}\right|_{\omega=\tilde\omega}(\omega-\tilde\omega)+\left.\frac{1}{2}\frac{\partial^2 \beta^2}{\partial \omega^2}\right|_{\omega=\tilde\omega}(\omega-\tilde\omega)^2 + \cdots.
\end{equation}
Let us denote $\tilde{\bb}=\beta(\tilde{\oo})$. Applying the inverse Fourier transform to (\ref{master_fourier}), the straightforward calculation yields 
\begin{eqnarray}
\frac{\ppy^2}{\ppy x^2}u(x,t) &+& {\mathrm{e}}^{i(\tilde\beta x-\tilde\omega t)} \sum_k \frac{1}{k!}\left.\frac{\partial^k \beta^2}{\partial \omega^k}\right|_{\omega=\tilde\omega} \left(i\frac{\partial}{\partial t}\right)^k U(x,t) \nonumber\\
&+& {\mathrm{e}}^{-(i\tilde\beta x-\tilde\omega t)} \sum_k
\frac{1}{k!}\left.\frac{\partial^k \beta^2}{\partial
    \omega^k}\right|_{\omega=-\tilde\omega}
\left(i\frac{\partial}{\partial t}\right)^k U^*(x,t) \nonumber\\
&=&
\chi_3 \frac{\ppy^2}{\ppy t^2}u^3(x,t). \label{main}
\end{eqnarray}
Since we assume that the signal is concentrated around a carrier wave , we now introduce a slow time $t_1$ by setting
\begin{equation}
t_0=t, \qquad t_1=\eps t.
\end{equation}
Then we can write (\ref{wave_noscales}) as 
\begin{equation}
u(x,t_0,t_1) = U({x},t_1)e^{i(\tilde{\bb}x-\tilde{\oo} t_0)} +
U^*({x},t_1)e^{-i(\tilde{\bb}x-\tilde{\oo}t_0)}, \label{wave}
\end{equation}
where $U(x,t_1)$ is a slowly varying function of $x$. Here, notice that we do not explicitly separate the scales in the evolution variable ${x}$ at this step. Now we use a small amplitude expansion of the function $u$
\begin{equation}
u = \eps u_0 + \eps^2 u_1 + \eps^3 u_2 + \cdots.
\end{equation}
Solving now (\ref{main}) order by order, we can determine the equation
for $U(x,t_1)$ which describes the slowly varying amplitude of the
electric field $u$. 

Without any loss of generality we assume $\chi_3=1$. We first
collect terms $\mathcal{O}(\eps)$  and find
\begin{displaymath}
\left(\frac{\ppy^2}{\ppy x^2} + \tilde{\bb}^2\right)u_0(x,t_0,t_1) = 0.
\end{displaymath}
Assuming the solution has the form of (\ref{wave}), we plug it into the above equation. Due to the ansatz (\ref{wave}), at the lowest order we find the amplitude of solution $u_0$ does not depend on $x$ and obtain
\begin{displaymath}
u_0(x,t_0, t_1) = A_0(t_1) e^{i(\tilde{\bb}x-\tilde{\oo}t_0)}
+A_0^*(t_1) e^{-i(\tilde{\bb}x-\tilde{\oo}t_0)}.
\end{displaymath}
Here, $A_0(t_1)$ can be explicitly determined from the initial condition for (\ref{master}). In fact, one can verify for the homogeneous equation $\displaystyle(\frac{\partial^2}{\partial x^2} + \tilde{\beta}^2)u = 0$, the amplitude of the solution does not depend on $x$ due to our ansatz. We will now see how the $x-$dependence of amplitude enters. At the second order, $\mathcal{O}(\eps^2)$, we
find
\begin{eqnarray}
\left(\frac{\ppy^2}{\ppy x^2} + \tilde{\bb}^2\right)u_1(x,t_0,t_1) &+& \left.\frac{d\bb^2}{d\oo}\right|_{\oo =
  \tilde{\oo}}\left(i\frac{d}{d t_1}
  A_0(t_1)\right)e^{i(\tilde{\bb}x-\tilde{\oo} t_0)} \nonumber\\ &+&
  \left.\frac{d\bb^2}{d\oo}\right|_{\oo =
  -\tilde{\oo}}\left(i\frac{d}{d t_1}
  A_0^*(t_1)\right)e^{-i(\tilde{\bb}x-\tilde{\oo} t_0)} = 0.
  \label{2nd}
\end{eqnarray}
We will write the solution of (\ref{2nd}) as the sum of a particular solution $P(x,t_0,t_1)$ and the general solution $A_1(t_1) e^{i(\tilde{\bb}x-\tilde{\oo} t_0)} +A_1^*(t_1)
e^{-i(\tilde{\bb}x-\tilde{\oo} t_0)}$ of the homogeneous equation, i.e., we write 
\begin{displaymath}
u_1(x,t_0,t_1) = A_1(t_1) e^{i(\tilde{\bb}x-\tilde{\oo} t_0)} +A_1^*(t_1)
e^{-i(\tilde{\bb}x-\tilde{\oo} t_0)}+ P(x,t_0,t_1),
\end{displaymath}
where $A_1(t_1)$ depends on the initial condition and we denote
$P(x,t_0,t_1)$ by the particular solution to equation
(\ref{2nd}). One of the simple ways to find this particular solution
for the given equation is by letting
\begin{displaymath}
P(x, t_0,t_1) = e^{i\tilde{\bb}x}a(t_0,t_1)x
+e^{-i\tilde{\bb}x}b(t_0,t_1)x,
\end{displaymath}
where $a(t_0,t_1),b(t_0,t_1)$ are to be determined later.  Plugging
this into equation (\ref{2nd}) yields
\begin{displaymath}
P(x,t_0,t_1) = -\tilde{\beta}'\left(\frac{dA_0(t_1)}{dt_1} x e^{i(\tilde{\bb}x-\tilde{\oo}
t_0)}+\frac{dA_0^*(t_1)}{dt_1} x e^{-i(\tilde{\bb}x-\tilde{\oo}
t_0)}\right).
\end{displaymath}
Then, we obtain the second order approximate solution
\begin{eqnarray*}
u^{(2)}(x,t_0,t_1) &=& \eps (u_0 + \eps u_1)\\
&=& \eps \left(A_0(t_1) + \eps {A_1(t_1)}- \eps \tilde{\bb}'\frac{dA_0(t_1)}{dt_1} x \right)
e^{i(\tilde{\bb}x-\tilde{\oo} t_0)} \\
&+& \text{complex conjugate}.
\end{eqnarray*}
Letting $\tilde{A}_0(t_1) = A_0(t_1) + \eps A_1(t_1)$, we find
\begin{eqnarray*}
u^{(2)}(x,t_0,t_1) 
&=& \eps \left(\tilde{A}_0(t_1) - \eps \tilde{\bb}'\frac{d \tilde{A}_0(t_1)}{dt_1} x \right)
e^{i(\tilde{\bb}x-\tilde{\oo} t_0)} + \mathcal{O}(\eps^3)\\
&+& \text{complex conjugate}.
\end{eqnarray*}
Since $u^{(2)}(x,t_0,t_1)$ needs to be an approximation valid up to
order $\mathcal{O}(\eps^2)$, the term $\mathcal{O}(\eps^3)$ can be
neglected. Hence, we have
\begin{eqnarray*}
u^{(2)}(x,t_0,t_1) 
&=& \eps \left(\tilde{A}_0(t_1) - \eps \tilde{\bb}'\frac{d \tilde{A}_0(t_1)}{dt_1} x \right)
e^{i(\tilde{\bb}x-\tilde{\oo} t_0)}\\
&+& \text{complex conjugate}.
\end{eqnarray*}
Here, we notice that a secular term appears, which corresponds to the term proportional to $x$. In other words,
this approximation is no longer valid when $x \sim
\mathcal{O}\left(\displaystyle{1}/{\eps}\right)$ or higher. 
In order to get rid of this secular term, we consider the term
$\tilde{A}_0(t_1) -
\eps\tilde{\bb}'\displaystyle\frac{d \tilde{A}_0(t_1)}{dt_1} x$ as the
order 1 Taylor expansion of some function $\Lambda(x,t_1)$ about
$x=0$. Thus, we need to find $\Lambda(x,t_1)$ which satisfies that
\begin{subequations}\label{s1y}
\begin{eqnarray}
\Lambda(x,t_1)|_{x = 0} &=& \tilde{A}_0(t_1),\label{1init}\\
\frac{\ppy\Lambda(x,t_1)}{\ppy x} &=& -\eps \tilde{\bb}'\frac{\ppy
  \Lambda(x,t_1)}{\ppy t_1}.\label{solv1}
\end{eqnarray}
\end{subequations}
This is {\it renormalization group equation}. Now the above form
of the equation motivates us to introduce a new scale $\eps x$.
Let us define $x_1 = \eps x$ then (\ref{s1y}) gives
\begin{subequations}
\begin{eqnarray}
\Lambda(x_1,t_1)|_{x_1 = 0} &=& \tilde{A}_0(t_1),\label{2init}\\
\frac{\ppy \Lambda(x_1,t_1)}{\ppy x_1} &=& - \tilde{\bb}'\frac{\ppy \Lambda(x_1,t_1)}{\ppy t_1}.\label{2ndf}
\end{eqnarray}
\end{subequations}
By solving (\ref{2ndf}) provided that the initial condition
(\ref{2init}) is satisfied, we can express $\tilde{A}_0(t_1) -
\tilde{\bb}'\displaystyle\frac{d \tilde{A}_0(t_1)}{dt_1} x_1 $ as the
Taylor expansion of $\Lambda(x_1,t_1)$ about
$x_1=0$. Then, we finally obtain the second order approximate
solution without a secular term
\begin{equation} \label{2ndorder_nlse_approx}
u^{(2)} =
\eps \Lambda(x_1,t_1) e^{i(\tilde{\bb}x -\tilde{\oo} t_0)}+\text{complex conjugate}.
\end{equation}
In fact, we can find the explicit form of $\Lambda(x_1,t_1)$, that is
$\Lambda(x_1,t_1) = \tilde{A}_0(\tilde{\bb}'x_1 - t_1)$. We notice that the solution of the RG equation $\Lambda(x_1,t_1)$ remains $\mathcal{O}(1)$. Thus, the second order approximate solution is valid over the time scales of interest.
In the same way we can find the contributions of higher orders. A
detailed calculation can be found in the Appendix. Since we want to
analyze the effect of nonlinearity, we need to explore the order
${\mathcal{O}}(\eps^3)$ where the effects of nonlinearity first
occur. The result of the application of the RG method yields
\begin{equation} \label{get_u_from_nlse}
u(x,t_0,t_1)=\eps \tilde\Sigma(x,t_1)e^{i(\tilde{\bb}x-\tilde{\oo} t_0)} + \eps \tilde\Sigma^*(x,t_1)e^{-i(\tilde{\bb}x-\tilde{\oo} t_0)}+{\mathcal{O}}(\eps^3)
\end{equation}
where the function $\tilde\Sigma(x,t_1)$ satisfies the NLSE of the form
\begin{eqnarray}
\frac{\ppy \tilde\Sigma(x,t_1)}{\ppy x} &=& -\eps\tilde{\bb}'\frac{\ppy \tilde\Sigma(x,t_1)}{\ppy
  t_1} \label{nlse_equation}\\ \nonumber && + \eps^2 i\left(-\frac{\tilde{\bb}''}{2}\frac{\ppy^2}{\ppy
 t_1^2}\tilde\Sigma(x,t_1) +
 \frac{3\chi_3}{2}\frac{\tilde{\oo}^2}{\tilde{\bb}}\tilde\Sigma(x,t_1)|\tilde\Sigma(x,t_1)|^2\right). 
\end{eqnarray}
In order to carry out a direct comparison between
(\ref{nl_max}) and (\ref{nlse_equation}) the only remaining
step is to specify in (\ref{nlse_equation}) the dispersion
$\beta(\omega)$.  We calculate this from the assumed form of
the susceptability, 
$\hat{\chi}(\omega)=-\chi_0/\omega^2$. In this case,
(\ref{disprelation}) yields
\begin{equation} \label{special_dispersion}
\tilde\beta'=\frac{\tilde\omega}{\sqrt{\omega^2-\chi_0}},\qquad \tilde\beta''=\frac{-\chi_0}{(\omega^2-\chi_0)^{3/2}}.
\end{equation}
As mentioned before, we expect the cubic nonlinear Schr{\"o}dinger
equation to be a good approximation for fairly broad pulses. For short
pulses the scaling is entirely different and, based on the results of
the previous section based on the linear case
(\ref{lin_ansatz_evolve}), it is reasonable to assume that for short
pulses we can approximate the solution of (\ref{nl_max}) by an ansatz
with the scaling of the form
\begin{equation} \label{nl_ansatz_evolve}
u(x,t) = \epsilon \cV_0(\tm,\eps x) +  \epsilon^2 \cV_1(\tm,\eps x) + ...
\end{equation}
into (\ref{nl_max}). The leading order, nontrivial behavior of
$ \cV_0$ is then given by \cite{schaefer-wayne:2003}
\begin{equation} \label{nl_sp_equation}
-2 \pp \partial_X \cV_0 = \chi_0 \cV_0 + \chi_3 \partial_{\phi\phi} (\cV_0)^3, \qquad X=\eps x.
\end{equation}
This nonlinear short-pulse equation describes the influence of the
nonlinear contribution of the polarization to the pulse. Note that
there is a slight difference in the nonlinear ansatz
(\ref{nl_ansatz_evolve}) from the linear result presented in
(\ref{lin_ansatz_evolve}), namely that the evolution variable is
$\epsilon x$ in contrast to $\epsilon t$. Since our system is weakly
dispersive, the pulses propagate on the leading order with speed one,
an approximation valid to ${\mathcal O}(1/\epsilon)$ in time should
also be valid to ${\mathcal O}(1/\epsilon)$ in space.

Eq. (\ref{nl_sp_equation}) can be derived by the
renormalization group (RG) method as well. Starting from
(\ref{nl_max})
\begin{displaymath}
u_{xx} = u_{tt} + \chi_0 u + \chi_3 (u^3)_{tt}.
\end{displaymath}
we first make a coordinate transform
\begin{equation}
u(x,t)=B\left(\frac{t-x}{\epsilon},x\right)
\end{equation}
which takes care of (a) the introduction of the time-scale for
ultra-short pulses, namely that the initial condition will depend on
$t/\epsilon$ and (b) a moving frame corresponding to a right-moving
pulse. The transformed equation for $B(\phi,x)$ is 
\begin{equation}
-\frac{2}{\epsilon}B_{\phi x} = \chi_0 B - B_{xx} + \chi_3 (B^3)_{\phi\phi}
\end{equation}
As we will see, the attempt to solve this equation by a standard,
weakly nonlinear perturbation expansion of the form
\begin{equation}
B(\phi,x) = \epsilon \Lambda_0(\phi,x) + \epsilon^2 \Lambda_1(\phi,x)
+ ...
\end{equation}
will lead to secular growth: At the leading order, we obtain that
$\Lambda_0$ does not depend on $x$. This reflects the fact that, in
our scaling, the leading order solution is the initial data traveling
to the right with the speed one. Or, in other words, the system we are
considering is {\em weakly} dispersive and nonlinear such that neither
dispersion nor nonlinearity enter the solution on the lowest order. At
the next order, however, both effects are present, since we find as a
solution for $\Lambda_1$
\begin{equation}
\Lambda_1 = -\frac{1}{2}\;x\;\int_{-\infty}^{\phi}\chi_0 \Lambda_0(\phi')+\chi_3 (\Lambda_0(\phi')^3)_{\phi \phi}\;d\phi'.
\end{equation}
The expansion for $B(\phi,x)$ is then
\begin{equation}
B(\phi,x)\approx \epsilon\left(\Lambda_0 - \frac{\epsilon}{2}\;x\;\int_{-\infty}^{\phi}\chi_0 \Lambda_0(\phi')+\chi_3 (\Lambda_0(\phi')^3)_{\phi \phi}\;d\phi'\right).
\end{equation}
In order to eliminate the growth of $\Lambda_1$ with respect to $x$, we look
for a transformation of the form
\begin{eqnarray}
\cV_0(\phi,x=0)&=&\Lambda_0(\phi),\\
\frac{\partial \cV_0}{\partial x} &=& -\frac{\epsilon}{2}\int_{-\infty}^{\phi}\chi_0 \Lambda_0(\phi')+\chi_3 (\Lambda_0(\phi')^3)_{\phi \phi}\;d\phi'
\end{eqnarray}
or, introducing a slow variable $X=\epsilon x$ and setting
$\cV_0=\cV_0(\phi,X)$, we find
\begin{equation}
-2\partial_{X}\partial_{\phi}\cV_0=\chi_0 \cV_0 + \chi_3\partial^2_{\phi}\left(\cV_0^3\right)
\end{equation}
which is exactly the short-pulse equation (\ref{nl_sp_equation}). In
this derivation we see again, how the scales of the evolution variable
$x$ appear naturally. Of course, we could have also used a
multiple-scale expansion of the form
\begin{displaymath}
B(\phi,x) = \epsilon \cV_0(\phi,x_0,x_1) + \epsilon^2 \cV_1(\phi,x_0,x_1) + ...
\end{displaymath}
which leads, as the reader can easily verify, to the same result.

We expect on the basis of the derivation of both approximate equations
and extensive numerical and experimental evidence that for broad
pulses the nonlinear Schr{\"o}dinger equation is an excellent
approximation but for ultra-short pulses, the nonlinear short-pulse
equation should be a more appropriate approximation than
NLSE. Intuitively this is clear by the scaling of the ansatz that was
used in both derivations: In order to derive (\ref{nl_sp_equation}) we
started from a pulse of the form ${\mathcal{U}}(t/\eps)$ whereas the
NLSE describes the envelope on a slow time scale $A(x,\eps t)$. On the
other hand, it is not clear how far we can push each of those
perturbations.  It would be interesting to compare
(\ref{nl_sp_equation}) and (\ref{nlse_equation}) analytically to the
solution of Maxwell's equations given by (\ref{nl_max}), but this is
an extremely complicated problem. Therefore, we approach this problem
numerically.

\section{Numerical comparison of the approximations to Maxwell's equations}

We perform the following (numerical) experiment: Consider Maxwell's
equation (\ref{nl_max}) with an initial data that corresponds to pulse
carried by a carrier wave:
\begin{equation} \label{nl_pulse}
u(x=0,t) = a\;{\mathrm{e}}^{-b^2t^2/2}\;\cos{\omega_0 t}.
\end{equation}
Here, we choose $x$ to be our evolution variable and choose
$u_x(x=0,t)$ such that our initial conditions correspond to a forward
traveling wave of the linear problem. The factor $a$ in
(\ref{nl_pulse}) corresponds to the amplitude of the pulse and the
parameter $b$ determines the pulse width. For the susceptibility
$\chi(\omega)$ we use in the numerical simulations the following form
\begin{equation}
\hat{\chi}(\omega)=-\frac{\chi_0}{\omega^2}\left({\mathcal{H}}(\omega-\omega_c)+{\mathcal{H}}(-\omega_c-\omega)\right)
\end{equation}
with ${\mathcal{H}}(x)=1$ for $x\leq 0$ and ${\mathcal{H}}(x)=0$
elsewhere. This accounts for the fact that the pulse cannot propagate
for very low frequencies. We assume that the amplitude of the pulse
decays to almost zero at $\omega_c$. This condition is
satisfied by an appropriate choice of the carrier frequency $\omega_0
> \omega_c$. First, we discuss the ``classical'' case for small $b$,
where the NLSE applies. To apply the NLSE model, we first extract out
of (\ref{nl_pulse}) the corresponding initial condition for the slowly
varying amplitude, compute the evolution of this initial data
according to (\ref{nlse_equation}) and then use
(\ref{get_u_from_nlse}) in order to reconstruct the electric field
$u(x_{\mathrm{end}},t)$. 
\begin{figure}[htb]
\centering
\includegraphics[width=0.425\textwidth, bb = 50 50 554 770, angle=-90]{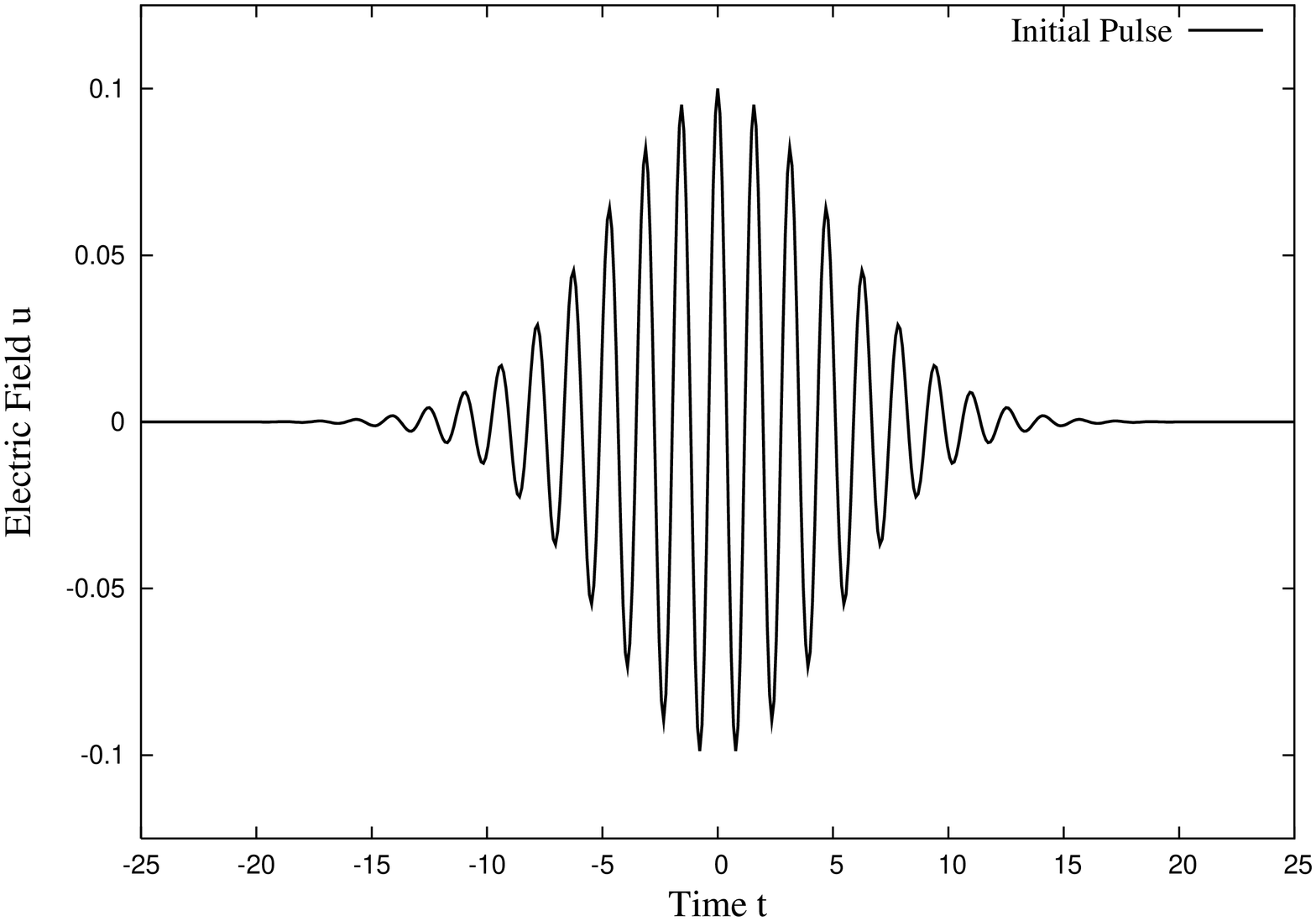}
\hfill
\includegraphics[width=0.425\textwidth, bb = 50 50 554 770, angle=-90]{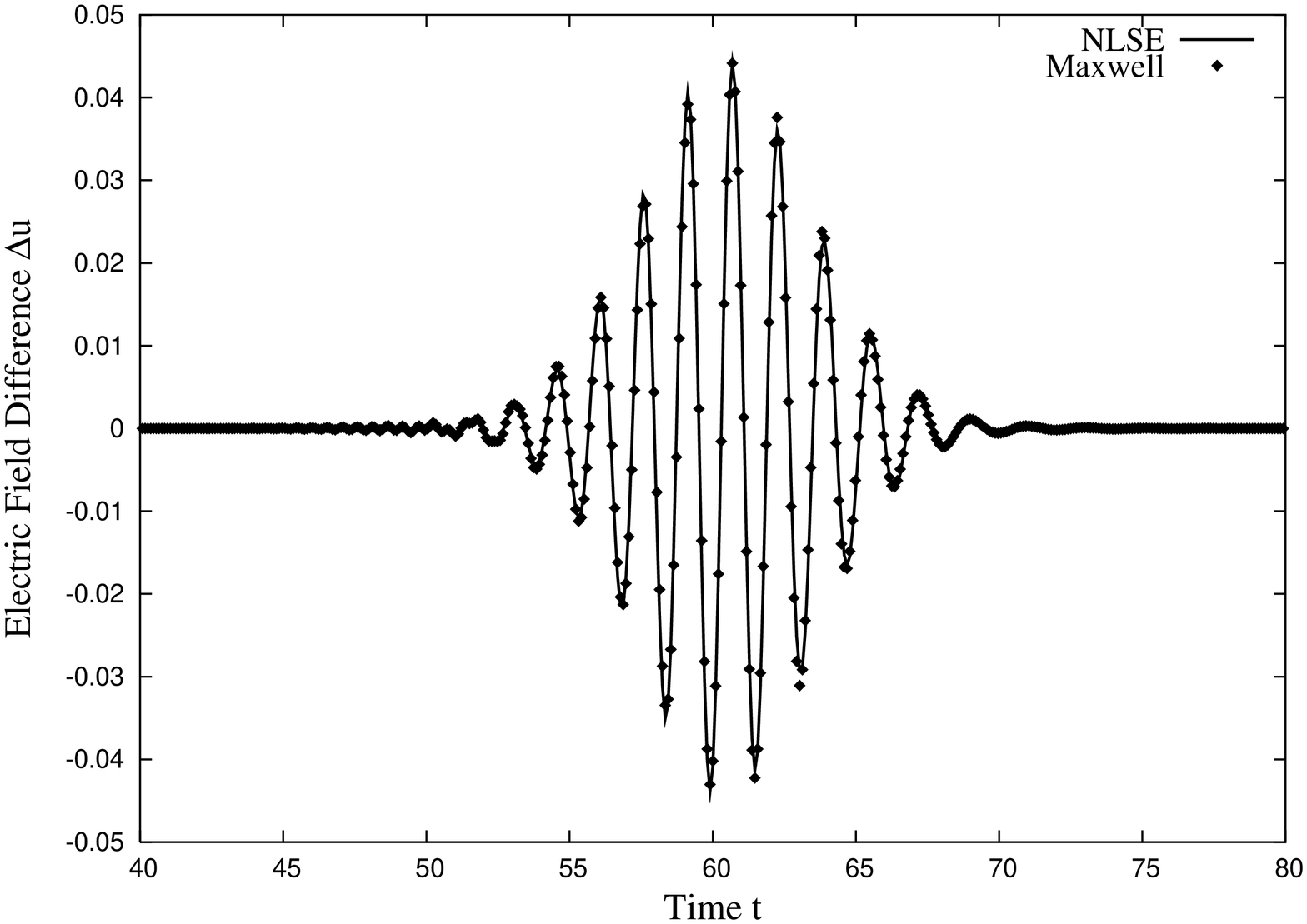}
\caption{{\protect\small Comparison of the solution of Maxwell's equation and the cubic nonlinear Schr{\"o}dinger equation. The figure above shows the initial pulse at $x=0$. The figure below compares the difference of nonlinear Maxwell's equation and the solution of the corresponding linear problem to the prediction of the cubic nonlinear Schr{\"o}dinger equation. The total propagation distance is $x_{\mathrm{end}}=50$. The parameters of the simulation are $a=0.1$, $b=0.2$, $\omega_c = 2.5$, $\omega_0=4$, $\chi_0=5$ and $\chi_3=0.5$. Eq. (\ref{special_dispersion}) yields $\tilde\beta'\approx 1.2$.}}
\label{fig:broad.ps}
\end{figure}
Figure \ref{fig:broad.ps} shows the typical result of the case where
the nonlinear Schr{\"o}dinger approximation holds. Due to the choice
of $b=0.2$, the width of the pulse is sufficiently large in comparison
to the period of the carrier frequency. Since we are interested in the
{\em nonlinear} evolution of the signal, we present the result of the
NLSE approximation in the following form: First, we propagate the
initial pulse in the linear Maxwell's equations by setting
$\chi_3=0$. This solution $u_{\mathrm{lin}}$ serves as reference data:
Now we propagate the same pulse in the corresponding nonlinear setting
and, after obtaining $u_{\mathrm{maxwell}}$, we compute the difference
from the linear solution
\begin{equation}
\Delta u_{\mathrm{maxwell}} = u_{\mathrm{maxwell}}-u_{\mathrm{lin}}.
\end{equation}
Then we find by solving the NLSE the corresponding $u_{\mathrm{nlse}}$
and compute
\begin{equation}
\Delta u_{\mathrm{nlse}} = u_{\mathrm{nlse}}-u_{\mathrm{lin}}.
\end{equation}
The question is now how well $\Delta u_{\mathrm{nlse}}$ approximates
$\Delta u_{\mathrm{maxwell}}$. As we can see from figure
\ref{fig:broad.ps}, in this case, the approximation is excellent. Let 
us now increase $b$ corresponding to making the pulse shorter.
\begin{figure}[htb]
\centering
\includegraphics[width=0.425\textwidth, bb = 50 50 554 770, angle=-90]{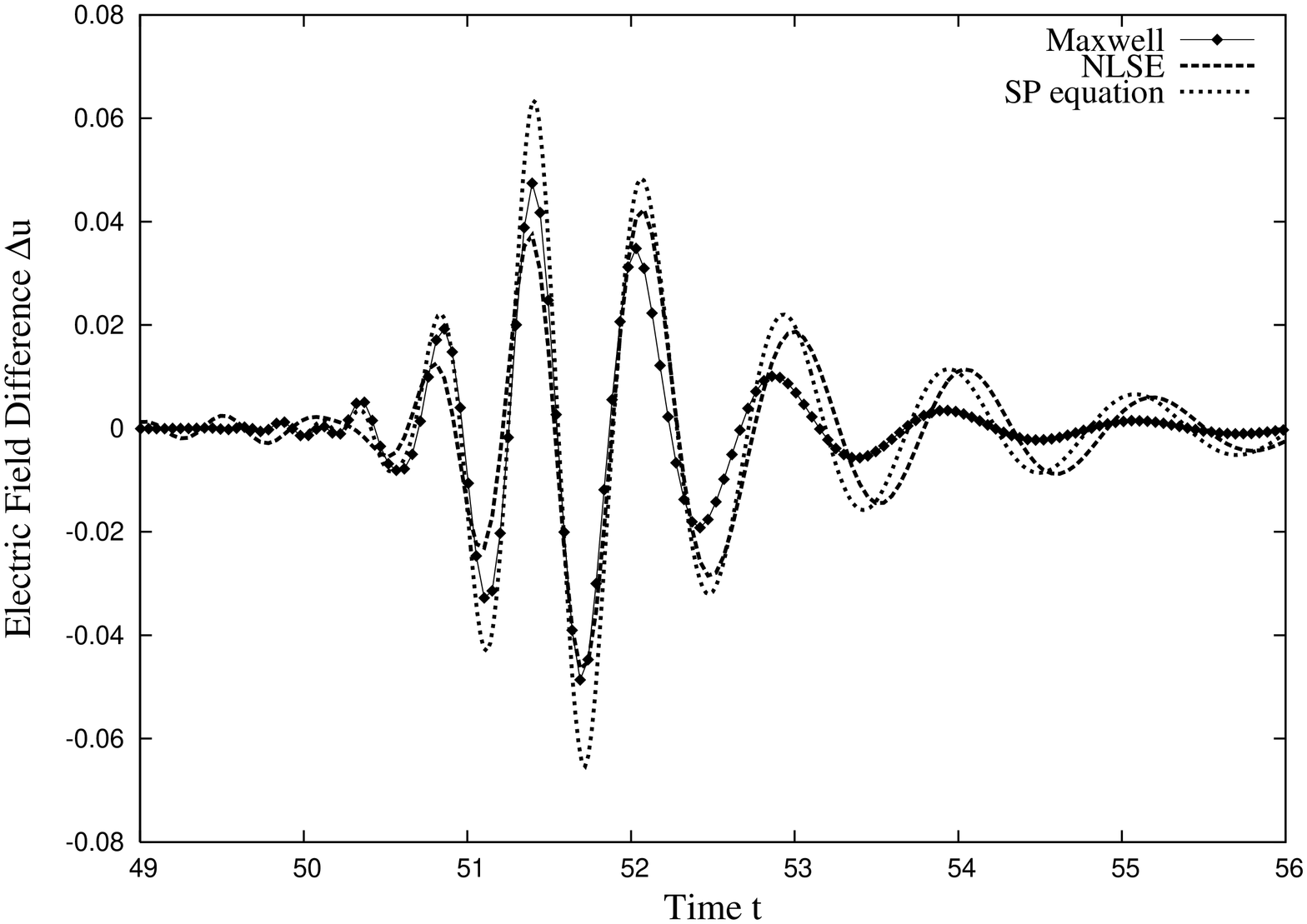}
\hfill
\includegraphics[width=0.425\textwidth, bb = 50 50 554 770, angle=-90]{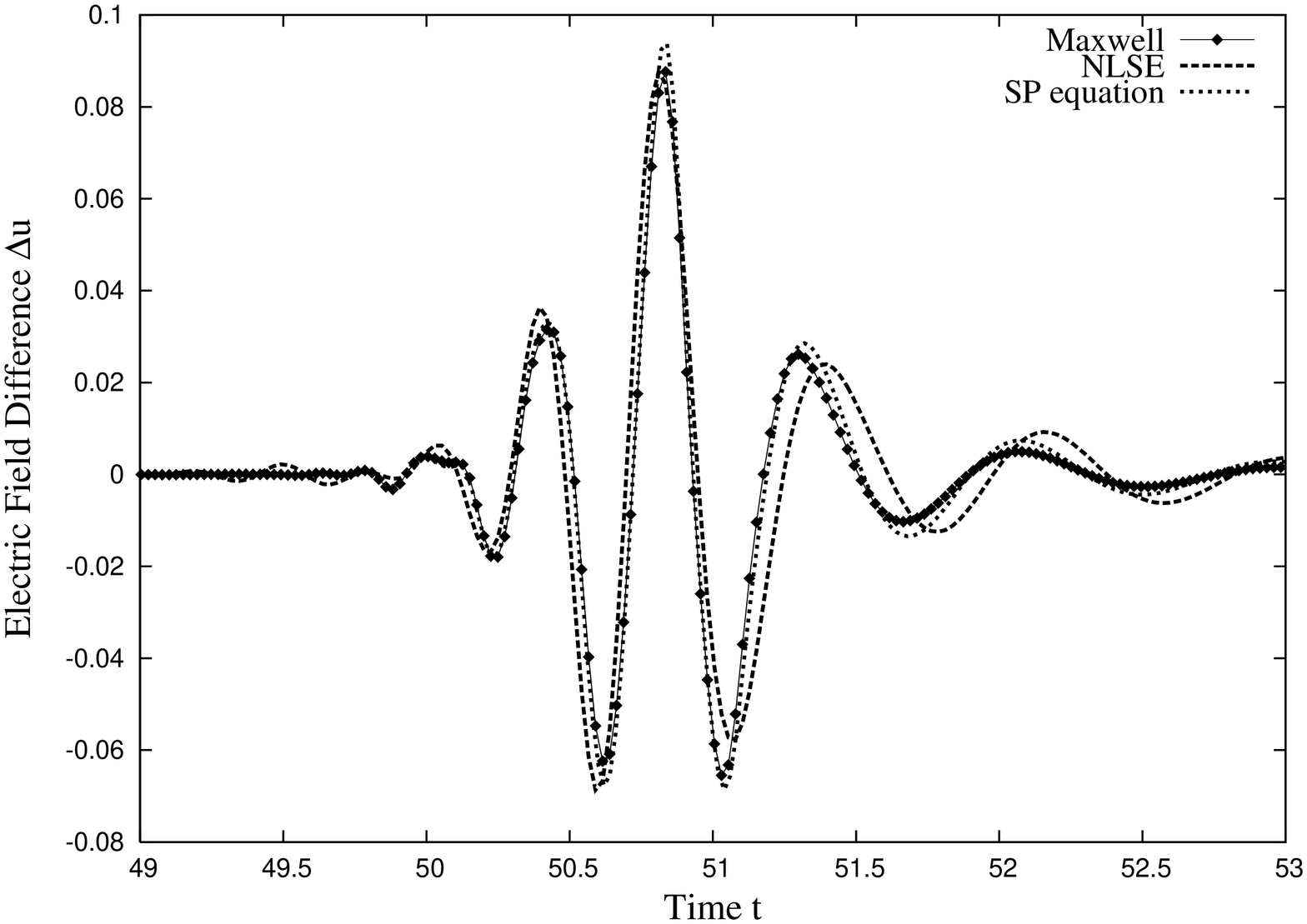}
\caption{{\protect\small Comparison of the solution of Maxwell's equation and the cubic nonlinear Schr{\"o}dinger equation and the short-pulse equation for a short pulse. Again, the figures compare the difference of nonlinear Maxwell's equation and the solution of the corresponding linear problem to the prediction of the cubic nonlinear Schr{\"o}dinger equation and the short-pulse equation. The parameters for this simulation are the same as in figure \ref{fig:broad.ps} with the exception of $b=2$, $\omega_0=6.5$ for the figure above and $b=3$, $\omega_0=13$ for the figure below.}}
\label{fig:medium.ps}
\end{figure}
Let's first choose $b=2$. Notice that this parameter already formally
violates the basic assumption about separation of time scales made in
the derivation of the NLSE. Therefore, from figure
\ref{fig:medium.ps}, it is surprising how well the NLSE still works.
In this parameter regime, the short-pulse equation is not better than
the NLSE since for $b=2$ the essential assumption about a $t/\eps$
dependence of the initial condition is not satisfied. It is possible
to extend the validity of NLSE to shorter pulses by incorporating
higher order terms. In the Appendix, we also give a derivation for the
next order that appears in the RG expansion. In this article, however,
we want to focus on the comparison between the leading order
approximation of Maxwell's equations and the ultra-short pulse
equation. Going to shorter pulses, e.g. for $b=3.0$, already, the
short-pulse equation starts to do a better job than the NLSE as we can
see from Figure
\ref{fig:medium.ps}. 
\begin{figure}[htb]
\centering
\includegraphics[width=0.425\textwidth, bb = 50 50 554 770, angle=-90]{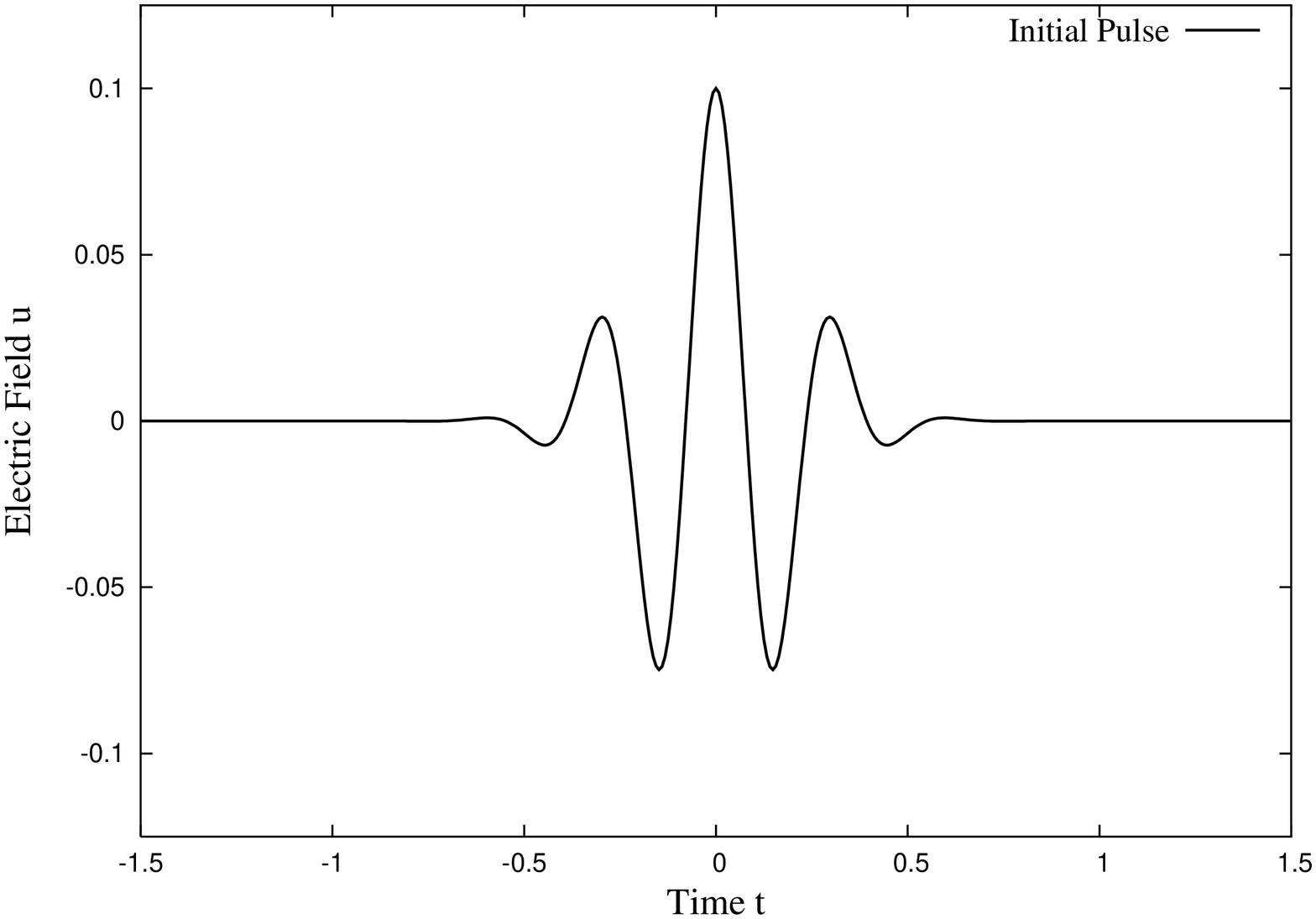}
\hfill
\includegraphics[width=0.425\textwidth, bb = 50 50 554 770, angle=-90]{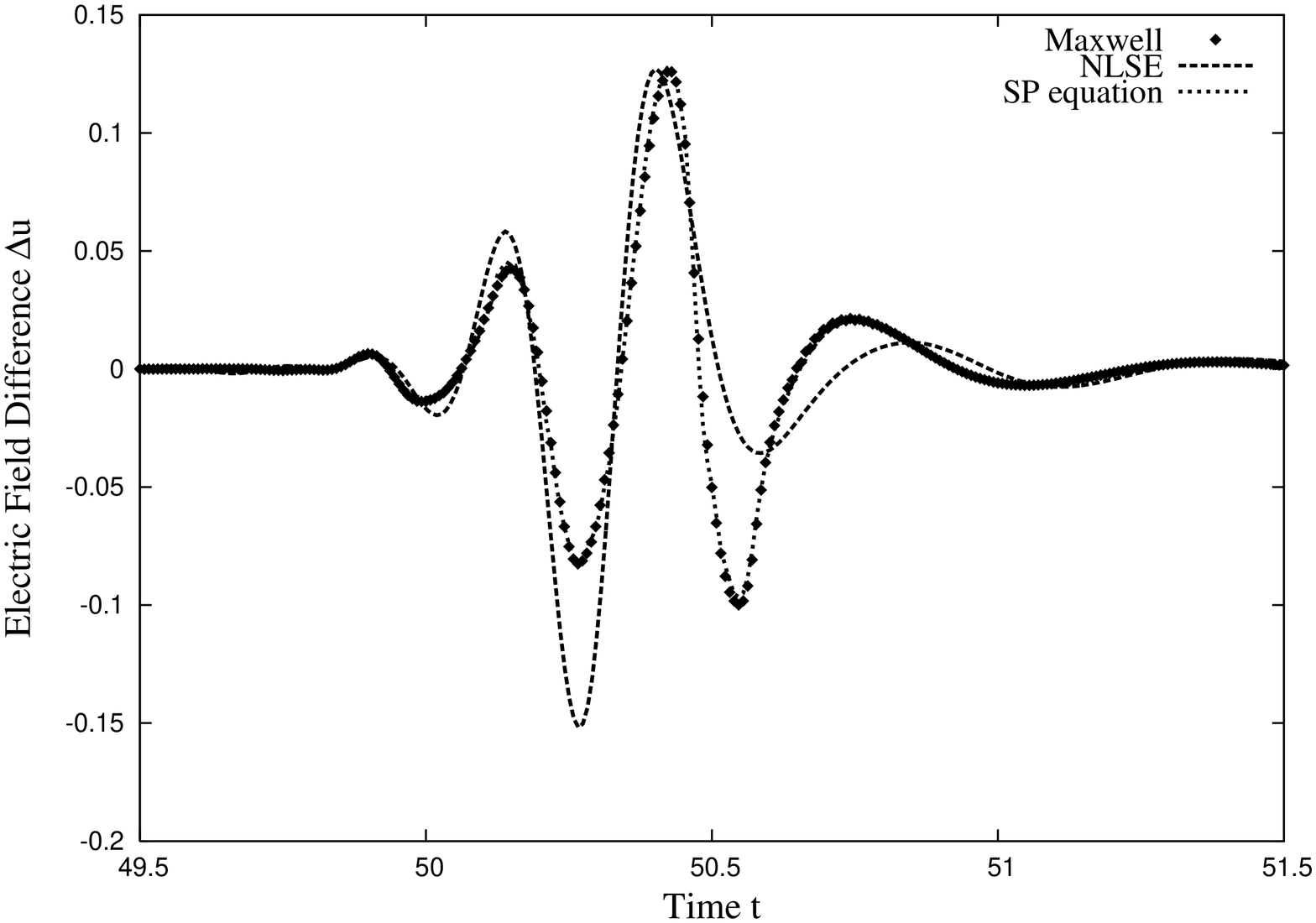}
\caption{{\protect\small Comparison of the solution of Maxwell's equation 
and the short-pulse equation and the NLSE for an ultra-short pulse. Again, 
the figure above shows the initial pulse at $x=0$ 
and the figure below compares 
the difference of nonlinear Maxwell's equation and the 
solution of the corresponding linear problem to the 
prediction of short-pulse equation and NLSE. The parameters 
for this simulation are the same 
as in figure \ref{fig:broad.ps} with the exception 
of $b=5$ and $\omega_0=20$.}}
\label{fig:short.ps}
\end{figure}
Setting $b=5$, we finally arrive to the domain of ultra-short
pulses. Here, the NLSE still predicts the rough shape of the pulse,
but does not give correct information about the pulse shape as we can
see from figure \ref{fig:short.ps}. On the other hand we can see now
that the short-pulse equation already in this chosen case of $b=5.0$
corresponding to $\eps=0.2$ provides an excellent approximation of
Maxwell's equation. Note that we propagated till
$x_{\mathrm{end}}=50\sim
{\mathcal{O}}\left(\frac{1}{\eps^2}\right)$. This numerical experiment
cannot substitute for a more thorough analytical investigation, but it
is an encouraging sign that, for ultra-short pulses,
(\ref{nl_sp_equation}) can be used in order to approximate the
solutions of (\ref{nl_max}). In further numerical studies we found
that even at propagation distances up to $x_{\mathrm{end}}=200$, the
short-pulse equation is a good approximation to Maxwell's equations.

\section{Conclusion}

In this paper we presented two main results: First we showed that if
we ignore nonlinear effects one can {\em rigorously} approximate the
evolution of a very short pulse injected into one end of an optical
fiber by a solution of the (linear) short-pulse equation
(\ref{pulse}). Second, in the case of nonlinear pulse propagation, we
have derived both the cubic nonlinear Schr{\"o}dinger equation and the
short-pulse equation directly from Maxwell's equations using
renormalization groups. We have shown numerically that the (nonlinear)
short-pulse equation gives an excellent approximation to the solution
of Maxwell's equation in the case of ultra-short pulses. The
analytical proof that those two solutions stay close, at least for an
evolution up to ${\mathcal{O}}(1/\eps)$ is a challenging task and
subject to future research.

\begin{appendix}
\section{Bounds on the roots of $Q$}\label{eigen}

We begin by proving the results in Lemmas \ref{s0} and \ref{s1} on the
large $p$ asymptotics of the roots of $Q$.  We give all the details of
the calculations needed to prove Lemma \ref{s0} and leave the very
similar details of the proof of Lemma \ref{s1} to the reader.  Begin
by defining $\sigma^0_{\pm}$ by
\begin{equation}
s^0_{\pm} = -\frac{\epsilon}{2}\left(\Gamma \pm \sqrt{\Gamma^2 -
4 \omega_0^2}\right) + \epsilon \sigma^0_{\pm}\ .
\end{equation}
Then $s^0_{\pm}$ is a root of $Q(s;p,\epsilon)$ if and only if
$\sigma^0_{\pm}$ is a solution of
\begin{eqnarray}\nonumber
&& g(\sigma^0_{\pm};p,\epsilon) = (\sigma^0_{\pm})^2 - \sigma^0_{\pm}
\left(\pm\sqrt{\Gamma^2 -
4 \omega_0^2}\right)  + \frac{\epsilon^2 \chi_0 
\left( -\frac{1}{2}\left(\Gamma \pm \sqrt{\Gamma^2 -\omega_0^2}
+ \sigma^0_{\pm}\right)\right)^2}{p^2} \\ \nonumber
&& \qquad   + \frac{\epsilon^2 \left((\sigma^0_{\pm})^2 - 
\sqrt{\Gamma^2 - 4 \omega_0^2}\right)\left(-\frac{1}{2}
\left(\Gamma \pm \sqrt{\Gamma^2 -
4 \omega_0^2}\right) + \epsilon \sigma^0_{\pm}\right)^2}{p^2} = 0\ .
\end{eqnarray}
If we let $\xi = p^{-2}$, and define $\tilde{q}(\sigma^0_{\pm};\xi,\epsilon)
= q(\sigma^0_{\pm};p,\epsilon)$ then we see that
$\tilde{q}(0;0,0)=0$ and $\partial_{\sigma} \tilde{q}(0;0,0)
= -\sqrt{\Gamma^2 -
4 \omega_0^2} \ne 0$, so the implicit function theorem implies that
there exists a smooth function $\sigma^0_{\pm}(\xi,\epsilon)$ such
that $\tilde{q}(\sigma^0_{\pm}(\xi,\epsilon);\xi,\epsilon) =0$.
Note that there exist constants
$A$, $B$, $C$, and $D$ which depend on 
$\Gamma$, $\omega_0$, and $\chi_0$, but not
on $\epsilon$ or $p$ such that if we rearrange the equation 
$\tilde{q}(\sigma^0_{\pm}(\xi,\epsilon);\xi,\epsilon) =0$,
(and replace $\xi$ by $p^{-2}$ it can be written as:
\begin{equation}\label{sigma_est}
|\sigma^0_{\pm}(\xi,\epsilon)| \le
(1 - A\epsilon^2/p^2)^{-1}\left( \frac{B \epsilon^2}{p^2} + 
C |\sigma^0_{\pm}(\xi,\epsilon)|^2 + 
\frac{D \epsilon^4}{p^2} |\sigma^0_{\pm}(\xi,\epsilon)|^4\right)\ .
\end{equation}
Since $\sigma^0_{\pm}(\xi=0,\epsilon=0)=0$, \reff{sigma_est}
immediately implies that there exists $C_0, C_1 > 0$ such
that for $p > C_0 \epsilon$, 
\begin{equation}
|\sigma^0_{\pm}(\xi,\epsilon)| \le \frac{C_1 \epsilon^2}{p^2}\ ,
\end{equation}
which completes the proof of Lemma \ref{s0}.
\qed

We now prove Lemma \ref{neg}  We first note that
if $p \ne 0$, $Q$ has no purely imaginary eigenvalues.
This follows by assuming that there exists such an
eigenvalue -- say $s = i x$, for $x \in \real$.
Inserting this into $Q$ and equating real and imaginary
parts we see that $x$ must satisfy
\begin{eqnarray}
(\epsilon^2 \omega_0^2 - x^2)(p^2-x^2) - \epsilon^2 x^2 
\chi_0 &=& 0 \\ \nonumber
\epsilon x \Gamma (p^2-x^2) &=& 0\ .
\end{eqnarray}
From the second of these equations we see that either
$x=0$ or $x= \pm p$. However, neither of these values
of $x$ solves the first equation and hence there are no
pure imaginary roots. Next note that from 
Lemmas \ref{s0} and \ref{s1} we know
that for $p$ sufficiently large, all four roots lie
in the left half plane.  But since the roots vary continuously
with $p$, the only way we could obtain a root with positive
real part was is one of the roots passed through the imaginary
axis.  We have just seen that there are no pure imaginary
roots for any non-zero value of $p$ and hence we never
have a root with positive real part.
\qed

Finally, we prove Lemma \ref{smallp}.  Note that if
we define $s=\epsilon \tilde{s}$ and $q=p/\epsilon$,
then $Q(s;p,\epsilon)=\tilde{Q}(\tilde{s},q,\epsilon)$,
with $\tilde{Q}(\tilde{s},q,\epsilon) =
(\tilde{s}^2 + \tilde{s}^2 \Gamma + \omega_0^2)(\tilde{s}^2+q^2)
+ \tilde{s}^2 \chi_0$, and the fact that the
roots of $Q$ can be written as $s^{0,1}_{\pm}(p)
= \epsilon \ts^{0,1}_{\pm}(p/\epsilon)$ follows.
Next note that for $q$ small, an easy perturbative
argument shows that $\tilde{Q}$ has a pair of complex
conjugate roots $\ts^0_{\pm}$ of size $\cO(1)$ (which correspond
to the roots $s^0_{\pm}$ of $Q$) and a pair of
complex conjugate roots $\ts^1_{\pm} = \pm i q + \cO(q^2)$.
Thus, Lemma \ref{smallp} will follow if we can show that for
all values of $q$ the roots of $\tilde{Q}$ are distinct.
Since the coefficients of $\tilde{Q}$ are real, the only
way it can have multiple roots is if there is a multiple
root on the (negative) real axis, or if there is a double
complex root $\ts$ and a second double root equal to the
complex conjugate of $\ts$.  We can immediately rule
out the first possibility by noting that
$$
\tilde{Q}(\tilde{s},q,\epsilon)
= \ts^2(\ts^2 + \ts \Gamma + \omega_0^2+ \chi_0)
+ q^2 (\ts^2 + \ts \Gamma + \omega_0^2)\ .
$$
But since we have assumed that $\Gamma^2 < 4 \omega_0^2$, 
$(\ts^2 + \ts \Gamma + \omega_0^2)>0$ and 
$(\ts^2 + \ts \Gamma + \omega_0^2+\chi_0)>0$ (for real
values of $\ts$) and hence
$\tilde{Q} > 0$ for all real values of $\ts$.
To rule out the possibility that the roots of $\tilde{Q}$
are of the form $\ts$ and $\overline{\ts}$ assume that
that one can factor $\tilde{Q} = (s-\ts)^2(s-\overline{\ts})^2$.
Expanding this expression and equating coefficients of like
powers of $s$ with the expression for $\tilde{Q}$ above
one finds a similar contradiction.
\qed

\section{Details on the derivation of NLSE by renormalization group method}

In this appendix we show how to solve for the higher orders
approximating the solution of (\ref{main}). Following the steps that
led us to (\ref{2ndorder_nlse_approx}), let us now find the third
order approximation. First, we need to collect $\mathcal{O}(\eps^3)$
terms. The usual way of obtaining these is simply plugging the
previous ansatz $u =
\eps u_0 + \eps^2 u_1 + \eps^3 u_2 + \cdots$ into (\ref{main}) and
collect appropriate terms. This will, however, lead to highly
complicated algebraic calculation. We now approach this problem by
making a different ansatz. Since we have already obtained the second
order approximation of the solution, we assume that 
\begin{displaymath}
u = u^{(2)} + \eps^3 u_2 + \cdots.
\end{displaymath}
For the nonlinear part of the equation (\ref{main}) we collect
$\mathcal{O}(\eps^3)$ terms which will give rise to secular growth. To
do this, we note that
\begin{eqnarray}
\frac{\ppy^2}{\ppy t^2}u(x,t)^3&=& -\epsilon^3{\mathrm{e}}^{3i(\tilde\beta x-\tilde\omega t)} \left(9\tilde\omega^2 + 6 \tilde{\oo} i \frac{\ppy}{\ppy t}\right)U(x,t)^3 \nonumber \\ 
                           &-& \epsilon^3{\mathrm{e}}^{i(\tilde\beta x-\tilde\omega t)} \left(3\tilde\omega^2 + 2\tilde{\oo}i \frac{\ppy}{\ppy t}\right) |U(x,t)|^2U(x,t) \nonumber \\
                           &-& \epsilon^3 {\mathrm{e}}^{-i(\tilde\beta x-\tilde\omega t)} \left(3\tilde\omega^2+ 2\tilde{\oo}i \frac{\ppy}{\ppy t}\right) |U(x,t)|^2U^*(x,t) \nonumber \\
                           &-& \epsilon^3{\mathrm{e}}^{-3i(\tilde\beta x-\tilde\omega t)} \left(9\tilde\omega^2+6 \tilde{\oo} i \frac{\ppy}{\ppy t}\right) U^*(x,t)^3 + \mathcal{O}(\epsilon^4). \label{nl}
\end{eqnarray}
Thus, at the third order ${\mathcal O}(\eps^3)$, we have
\begin{eqnarray*}
\left(\frac{\ppy^2}{\ppy x^2}\right. &+& \left.\tilde{\bb}^2\right)u_2 + \mathcal{F}_0({\mathrm{e}}^{\pm 3i(\tilde\beta x-\tilde\omega t_0)})\\
&+&\left(-\tilde{\bb} \tilde{\bb}'' \frac{\ppy^2}{\ppy t_1^2}
\Lambda(x_1,t_1)+ 3\tilde{\oo}^2
\Lambda(x_1,t_1)|\Lambda(x_1,t_1)|^2\right)e^{i(\tilde{\bb}x-\tilde{\oo} t_0)}\\
&+&\left(-\tilde{\bb} \tilde{\bb}''\frac{\ppy^2}{\ppy t_1^2}
\Lambda^*(x_1,t_1)+ 3\tilde{\oo}^2
\Lambda^*(x_1,t_1)|\Lambda(x_1,t_1)|^2\right)e^{-i(\tilde{\bb}x-\tilde{\oo} t_0)}
=0,
\end{eqnarray*}
where $\mathcal{F}_0({\mathrm{e}}^{\pm 3i(\tilde\beta x-\tilde\omega t_0)})$ contains the terms proportional to ${\mathrm{e}}^{\pm 3i(\tilde\beta x-\tilde\omega t_0)}$, and $\Lambda(x,t_1)$ is from (\ref{2ndorder_nlse_approx}).
Note that the terms that will lead to resonances (secular growth) are
the terms proportional to ${\mathrm{e}}^{\pm i(\tilde\beta
x-\tilde\omega t)}$.
Solving the above differential equation, we find 
$u_2 = A_2(t_1) e^{i(\tilde{\bb}x-\tilde{\oo}t_0)} + P(x,t_0,t_1) + \text{complex conjugate},$ where $A_2(t_1)$ is a function depending on the initial condition, and
\begin{eqnarray*} 
P(x,t_0,t_1) &=& -ix\left(\frac{\tilde{\bb}''}{2}\frac{\ppy^2}{\ppy
 t_1^2}\Lambda(x_1,t_1) - \frac{3}{2}\frac{\tilde{\oo}^2}{\tilde{\bb}}\Lambda(x_1,t_1)|\Lambda(x_1,t_1)|^2\right)e^{i(\tilde{\bb}x-\tilde{\oo} t_0)}\\
&+& \tilde{\mathcal{F}}_0(\mathrm{e}^{\pm 3i(\tilde\beta x-\tilde\omega t_0)}),
\end{eqnarray*}
where $\tilde{\mathcal{F}}_0(\mathrm{e}^{\pm 3i(\tilde\beta x-\tilde\omega t_0)})$ contains the terms proportional to ${\mathrm{e}}^{\pm 3i(\tilde\beta x-\tilde\omega t_0)}$. We can solve explicitly for the coefficients for these terms, and find these remain order 1 at all times. 
Now, the third order approximate solution is 
\begin{align*}
u^{(3)}& = \eps\Lambda(x_1,t_1) e^{i(\tilde{\bb}x-\tilde{\oo} t_0)} + \epsilon^3\tilde{\mathcal{F}}_0(\mathrm{e}^{\pm 3i(\tilde\beta x-\tilde\omega t_0)})\\
& + \eps^3\left(A_2(t_1)- ix\left(\frac{\tilde{\bb}''}{2}\frac{\ppy^2}{\ppy
 t_1^2}\Lambda(x_1,t_1) -
 \frac{3}{2}\frac{\tilde{\oo}^2}{\tilde{\bb}}\Lambda(x_1,t_1)|\Lambda(x_1,t_1)|^2\right)\right)
 e^{i(\tilde{\bb}x-\tilde{\oo}t_0)}\\
&+ \text{complex conjugate}.
\end{align*}
%
%
%
Here, to get rid of the secular terms, we need to find
$\Sigma(x,t_1)$ satisfying that
\begin{subequations}\label{m.4th}
\begin{eqnarray}
\Sigma(x,t_1)\mid_{x=0} &=& \Lambda(x_1,t_1)+\eps^2 {A_2}(t_1),\\
\frac{\ppy \Sigma(x,t_1)}{\ppy x} &=& \eps^2 i\left(-\frac{\tilde{\bb}''}{2}\frac{\ppy^2}{\ppy
 t_1^2}\Sigma(x,t_1) +
 \frac{3}{2}\frac{\tilde{\oo}^2}{\tilde{\bb}}\Sigma(x,t_1)|\Sigma(x,t_1)|^2\right). 
\end{eqnarray}
\end{subequations}
This leads us to introduce a new independent scale $x_2 = \eps^2 x$ in addition to $x_1 = \eps x$.
%
%
Therefore, from (\ref{m.4th}), we finally obtain
\begin{subequations}
\begin{eqnarray}
\Sigma\mid_{x_2=0} &=& \Lambda(x_1,t_1)+\eps^2 {A_2}(t_1),\\
\frac{\ppy \Sigma}{\ppy x_2} &=&
i\left(-\frac{\tilde{\bb}''}{2}\frac{\ppy^2}{\ppy t_1^2}\Sigma +
\frac{3}{2}\frac{\tilde{\oo}^2}{\tilde{\bb}}\Sigma|\Sigma|^2\right)\label{nlse}.
\end{eqnarray}
\end{subequations}
The equation (\ref{nlse}) is the {\it cubic nonlinear Schr\"odinger
equation}. We notice that the function $\Sigma$ now depends on
$x_1,x_2, t_1$, i.e., $\Sigma = \Sigma(x_1,x_2,t_1)$. From here,
together with (\ref{2ndf}), we obtain directly (\ref{nlse_equation})
by setting $\tilde\Sigma(x,t_1)=\Sigma(x_1,x_2,t_1)$. Following the
similar steps, we can extend the results to the higher order,
$\mathcal{O}(\eps^4)$ approximation. First, we assume the ansatz, $u =
u^{(3)} + \eps^4 u_3 + \cdots$ and recall (\ref{nl}). Recall ${\ppy t}
= \eps {\ppy t_1}$. Then, at the order $\mathcal{O}(\eps^4)$, we find
\begin{eqnarray*}
\left(\frac{\ppy^2}{\ppy x^2} + \tilde{\bb}^2 \right)u_3 &+&
e^{i(\tilde{\bb}x-\tilde{\oo} t_0)}\left(\frac{1}{3}\tilde{\bb}\tilde{\bb}'''(-i) \frac{\ppy^3}{\ppy t_1^3}
\Sigma\right.+\left.
i\left(6\tilde{\oo}-3\frac{\tilde{\bb}' \tilde{\oo}^2}{\tilde{\bb}}\right)\frac{\ppy}{\ppy t_1} \left(\Sigma|\Sigma|^2\right)\right)\\
&+&e^{-i(\tilde{\bb}x-\tilde{\oo} t_0)}\left(\frac{1}{3}\tilde{\bb}\tilde{\bb}'''i \frac{\ppy^3}{\ppy t_1^3}
\Sigma^*\right.
-\left.i\left(6\tilde{\oo}-3\frac{\tilde{\bb}' \tilde{\oo}^2}{\tilde{\bb}}\right) \frac{\ppy}{\ppy t_1} \left(\Sigma|\Sigma|^2\right)\right)\\
&+& \mathcal{F}_{\mathrm{nres}}=0,
\end{eqnarray*}
where $\mathcal{F}_{\mathrm{nres}}$ contains the non-resonant terms,
e.g. terms proportional to $e^{\pm 3i(\tilde{\bb}x-\tilde{\oo} t_0)}$,
$e^{\pm 4i(\tilde{\bb}x-\tilde{\oo} t_0)}$ etc. that are generated
when we insert $u_2$ back into the nonlinearity. Solving the above
differential equation, we find
\begin{eqnarray*}
u_3 &=& {A_3}(t_1)e^{i(\tilde{\bb}x-\tilde{\oo} t_0)} + \tilde{\mathcal{F}}_{\mathrm{nres}}\\
& -& \frac{1}{2i\tilde{\bb}}x e^{i(\tilde{\bb}x-\tilde{\oo} t_0)}
\left(\frac{1}{3}\tilde{\bb}\tilde{\bb}'''(-i) \frac{\ppy^3}{\ppy t_1^3}\Sigma\right.
+i \left.\left(6\tilde{\oo}-3\frac{\tilde{\bb}' \tilde{\oo}^2}{\tilde{\bb}}\right) \frac{\ppy}{\ppy t_1} \left(\Sigma|\Sigma|^2\right)\right)\\
&+& \text{complex conjugate},
\end{eqnarray*}
where $A_3(t_1)$ is a function depending on the initial condition, and
$\tilde{\mathcal{F}}_{\mathrm{nres}}$ contains the terms proportional
to higher powers of $e^{i(\tilde{\bb}x-\tilde{\oo} t_0)}$ whose
coefficients remain order 1 regardless of $x$.  Hence, we obtain the
fourth order approximate solution,
\begin{eqnarray*}
u^{(4)}&=& \eps \Sigma(x_1,x_2,t_1)e^{i(\tilde{\bb}x-\tilde{\oo} t_0)} + \epsilon^3\tilde{\mathcal{F}}_0(e^{\pm 3i(\tilde{\bb}x-\tilde{\oo} t_0)}) + \epsilon^4 \tilde{\mathcal{F}}_{\mathrm{nres}}\\
&+& \eps^4e^{i(\tilde{\bb}x-\tilde{\oo} t_0)}\left(A_3(t_1) + x \left(\frac{1}{6}\tilde{\bb}\tilde{\bb}'''\frac{\ppy^3}{\ppy t_1^3}
\Sigma(x_1,x_2,t_1)\right.\right.\\
&-&\left.\left.\frac{1}{2\tilde{\bb}}\left(6\tilde{\oo}-3\frac{\tilde{\bb}' \tilde{\oo}^2}{\tilde{\bb}}\right) \frac{\ppy}{\ppy t_1} \left(\Sigma(x_1,x_2,t_1)|\Sigma(x_1,x_2,t_1)|^2\right)\right)\right)\\
&+& \text{complex conjugate}.
\end{eqnarray*}
%
In order to eliminate the secular term, we need to find $V(x,t_1)$ satisfying
\begin{subequations}
\begin{eqnarray}
V(x,t_1)\mid_{x=0} &=& \Sigma(x_1,x_2,t_1) + \eps^3 A_3(t_1),\\
\frac{\ppy V}{\ppy x} &=& 
\eps^3\left(\frac{1}{6}\tilde{\bb}''' \frac{\ppy^3}{\ppy t_1^3}
V -\frac{1}{2\tilde{\bb}}\left(6\tilde{\oo}-3\frac{\tilde{\bb}' \tilde{\oo}^2}{\tilde{\bb}}\right) \frac{\ppy}{\ppy t_1} \left(V|V|^2\right)\right). \label{4theq}
\end{eqnarray}
\end{subequations}
This makes it natural to introduce a new scale $x_3 = \eps^3 x$. From (\ref{4theq}) it follows that
\begin{align*}
\frac{\ppy V}{\ppy x_3} = 
\frac{1}{6} \tilde{\bb}'''\frac{\ppy^3}{\ppy t_1^3}
V
-\frac{1}{2\tilde{\bb}}\left(6\tilde{\oo}-3\frac{\tilde{\bb}' \tilde{\oo}^2}{\tilde{\bb}}\right) \frac{\ppy}{\ppy t_1}\left(V|V|^2\right).
\end{align*}
Since $\displaystyle \frac{\ppy}{\ppy t_1} V|V|^2$ includes $|V|^2
\displaystyle\frac{\ppy}{\ppy t_1}V$ and
$V\displaystyle\frac{\ppy}{\ppy t_1}|V|^2$, we see that the Raman
scattering and the self-steepening terms appear at this high-order
approximation.

\end{appendix}

\section*{Acknowledgments}

This material is based upon work supported in part by the National Science
Foundation under Grants No.~0073923 and DMS-0103915.  TS and
CEW also wish to thank Jeffrey Rauch for stimulating correspondence
about the propagation of short pulses.

\bibliography{master}

\end{document}